\documentclass[final,seceqn]{elsart1p}
\usepackage{graphicx,epsf,bm,amssymb,amsbsy}
\journal{Nuclear Physics B}
\newcommand{\ve}{\varepsilon}
\newcommand{\be}{\begin{equation}}
\newcommand{\ee}{\end{equation}}
\newcommand{\bea}{\begin{eqnarray}}
\newcommand{\eea}{\end{eqnarray}}
\newcommand {\mm}[1]{\quad\mbox{#1}\quad}
\newcommand {\mmm}[1]{\quad\quad\mbox{#1}\quad\quad}
\newcommand {\lr}[1]{\left({#1}\right)}
\newcommand {\lrs}[1]{\left[{#1}\right]}
\renewcommand{\Re}{\mathrm{Re}}
\begin{document}
\begin{frontmatter}
\title{Lifshitz-point correlation length exponents from the
large-$n$ expansion}
\author[ICMP,FB]{M. A. Shpot},
\author[SP,FB]{Yu. M. Pis'mak}
\address[ICMP]{Institute for Condensed Matter Physics, 79011 Lviv, Ukraine}
\address[FB]{Fakult\"at f\"ur Physik, Universit\"at
Duisburg-Essen, D-47048 Duisburg, Federal Republic of Germany}
\address[SP]{State University of Sankt-Petersburg,
198504 Sankt-Petersburg, Russia}
\begin{abstract}
The large-$n$ expansion is applied to the calculation of thermal critical
exponents describing the critical behavior of spatially anisotropic
$d$-dimensional systems at $m$-axial Lifshitz points.
We derive the leading non-trivial $1/n$ correction for the
perpendicular correlation-length exponent $\nu_{L2}$ and hence several
related thermal exponents to order $O(1/n)$.
The results are consistent with known large-$n$ expansions for
$d$-dimensional critical points and isotropic Lifshitz points, as well as
with the second-order epsilon expansion about the upper critical dimension
$d^*=4+m/2$
for generic $m\in[0,d]$. Analytical results are given for the special case
$d=4$, $m=1$. For uniaxial Lifshitz points
in three dimensions, $1/n$ coefficients are calculated numerically.
The estimates of critical exponents at $d=3$, $m=1$ and $n=3$
are discussed.
\end{abstract}
\end{frontmatter}
\section{Introduction}

The (multi)critical behavior of strongly anisotropic systems
in the vicinity of a Lifshitz point (LP) \cite{HLS75,Hor80} is given by the
Landau-Ginzburg-Wilson (LGW) Hamiltonian
\be\label{EHI}
\mathcal H[\bm{\phi}]=\int d^dx\left[
\frac{\tau_0}{2}|\bm{\phi}|^2+\frac{1}{2}(\nabla_{\perp}\bm{\phi})^2
+\frac{\rho_0}{2}(\nabla_{\|}\bm{\phi})^2+
\frac{\sigma_0}{2}(\Delta_{\|}\bm{\phi})^2
+\frac{u_0}{4!}|\bm\phi|^4\right]
\ee
where $\bm \phi=\bm \phi(\bm x)$ is a classical $n$-vector
order-parameter field,
and the interaction is given by a standard $O(n)$ symmetric $\phi^4$ term.

Important is that physical properties along different spatial directions
essentially differ and this difference cannot be removed by simple
rescalings of the theory.\footnote{For discussions on this issue
see \cite[Ch. 4.2]{SZ95}, \cite{Dohm08}, \cite[pp. 15-17]{DC09}, \cite[pp. 13-14]{BDS10}, and
\cite{FAlt11} in the context of Lorentz violating theories.}
Accordingly, the $d$-dimensional coordinate space is split into
two Euclidean subspaces $\mathbb R^{d-m}$ and $\mathbb R^m$.
Each position vector $\bm x\in\mathbb R^{d-m}\times\mathbb R^m$ has a
$(d-m)$-dimensional "perpendicular" component $\bm x_{\perp}$ and an
$m$-dimensional "parallel" one, $\bm x_\|$.
The gradient operators $\nabla_{\perp}$ and $\nabla_\|$ act
in $\mathbb R^{d-m}$ and $\mathbb R^m$, respectively, and
$\Delta_\|\equiv\nabla_\|^2$ is the Laplacian associated
with parallel directions.
The term $(\Delta_{\|}\bm{\phi})^2$ is
fully rotationally invariant in $\mathbb R^m$.
Generically, less symmetric fourth-order derivative terms are
allowed to contribute into (\ref{EHI}) thus introducing
an additional anisotropy in $\mathbb R^m$ if $m>1$ \cite{DSZ03}.
These can be written as a linear combination
$w_aT_{ijkl}^a(\nabla_{\|,i}\nabla_{\|,j}\bm\phi)(\nabla_{\|,k}\nabla_{\|,l}\bm\phi)$
where $T$ are totally symmetric fourth-rank tensors compatible with the symmetry
of the system.

When the number of anisotropy axes $m$ shrinks to zero we
retrieve an isotropic $\phi^4_d$ theory with the upper critical dimension
$d^*=4$ and the usual critical-point (CP) behavior.
Another marginal situation is when $m$ extends to $d$.
The resulting $\phi^4_{m=d}$ theory with $d^*=8$ corresponds to the
isotropic LP \cite{HLS75,HLS75n,DS02}.
For generic $m\in[0,d]$, there is a line of upper critical dimensions
$d^*=d^*(m)=4+m/2$.

The $\phi^4$ coupling $u_0$, as well as the anisotropy parameter
$\sigma_0$, are positive constants, whereas $\tau_0$ and $\rho_0$
are allowed to change sign.
As usual, $\tau_0$ is a linear function of temperature, while
$\rho_0$ is related to an additional "non-ordering" field
specific for the underlying physical system (for a review see \cite{Die02})
and controls the crossover from LP.

In order to reach the LP, the parameters $\tau_0$ and $\rho_0$ have to be tuned to
their special values $\tau_0^{LP}$ and $\rho_0^{LP}$.
In the Landau approximation, $\tau_0^{LP}=\rho_0^{LP}=0$.
Similarly, in the renormalization-group theory the both
renormalized counterparts of $\tau_0$ and $\rho_0$ vanish at the LP: $\tau=\rho=0$.

It is demanded that the terms
$(\nabla_{\perp}\bm{\phi})^2$ and $\sigma_0(\Delta_{\|}\bm{\phi})^2$
scale in the same way. Hence,
in a cutoff-regularized theory, integrations over perpendicular and parallel
momenta $\bm k_{\perp}$ and $\bm k_\|$ have to be restricted asymmetrically
via $|\bm k_{\perp}|\le\Lambda$ and $|\bm k_\||\le\sigma_0^{-1/4}\Lambda^{1/2}$.
For large $\Lambda$, the shifts $\tau_0^{LP}$ and $\rho_0^{LP}$ behave as $\propto\Lambda^2$
and $\propto\Lambda$, respectively, while in dimensional regularization
(see e. g. \cite{Collins,ZJ89}) they vanish.

A simple dimensional analysis (cf. \cite{MC98,DS00,Vis09}) implies that
$$
[\tau_0]=-2[x_{\perp}],\; [x_\|]=\frac{1}{2}[x_{\perp}]+\frac{1}{4}[\sigma_0],\;
[\phi]=-\frac{1}{2}[d^dx]+[x_{\perp}]=\frac{1}{2}\big(d{-}\frac{m}{2}{-}2\big)[\mu]
{-}\frac{m}{8}[\sigma_0],
$$
$$
[\rho_0]=2[x_\|/x_{\perp}]=-[x_{\perp}]+\frac{1}{2}[\sigma_0],\quad
[u_0]=[\mu^\ve]+\frac{m}{4}[\sigma_0],\qquad\mbox{where}\qquad [\mu]=-[x_{\perp}],
$$
$\mu$ being an arbitrary momentum scale.
The deviation from the upper critical dimension $\ve$ is given by
$\ve\equiv d^*(m)-d=4+m/2-d$.
The (dimensionful) parameter $\sigma_0^{1/4}$ provides the scale for "measuring"
distances in the parallel subspace.
Its classical dimension is given by $[\sigma_0^{1/4}]=[x_\|/x_{\perp}^{1/2}]$.
The dimensionless combination $v\equiv\sigma_0^{-1/4}x_\|/x_{\perp}^{1/2}$ shows up
as the argument in the scaling function of the free propagator written in the scaling form
$G^{(0)}(x)=x_{\perp}^{-2+\ve}\sigma_0^{-m/4}\Phi(v)$.\footnote{
For general $m$ and $d$, the function $\Phi(v)$ has been derived in \cite{DS00,SD01}.
Its explicit expressions in different special cases can be found in
\cite{SG78,MC99,SD01,SPD05,AnsHal07}.}
The effective expansion parameter $\bar u_0\equiv u_0\sigma_0^{-m/4}$
has the usual $\mu$-dimension $[\mu^\ve]$ and becomes marginal at $d=d^*(m)$.
Similarly, the scaled field $\bm\phi(\bm x)\sigma_0^{m/8}$ becomes dimensionless at
the lower critical dimension \cite{GS78,Die02} $d_\ell(m)=2+m/2$
appropriate for the models with continuous $O(n)$ symmetry in the case $n>1$.

Provided that $d_\ell(m)<d<d^*(m)$, $0<m<d$, and the LP exists (see \cite{Die02,SPD05}),
the full two-point correlation function of the interacting theory obeys anisotropic scaling via
\be\label{VAS}
G(\bm x)=b^{2\Delta_\phi}G\lr{b\,\bm x_\perp,b^\theta\bm x_\|},\mm{where}
\Delta_\phi=\frac{1}{2}\left(d-m+\theta m-2+\eta_{L2}\right)
\ee
is the scaling dimension of the field, and all non-universal metric factors are omitted.
The anomalous field dimension $\eta_{L2}$ is similar to the usual Fisher exponent
$\eta$ of the pair correlation function at CP.
The key difference of (\ref{VAS}) from conventional CP scaling forms
is in that the distances along perpendicular and
parallel directions are rescaled with distinct scale factors: Here the anisotropy index
$\theta$ is different from $1$.
We stress also that $\theta$ deviates from its classical value $\theta_0=1/2$.
The difference, $\theta_1\equiv\theta-1/2$, is of order $O(\ve^2)$ in the epsilon expansion
\cite{DS00} and $O(1/n)$ in the large-$n$ expansion \cite{SPD05}.
Its non-zero value is provided by the non-trivial renormalization of the parameter $\sigma_0$.

We would like to recall that according to the definition of \cite{HLS75}
(see also reviews in \cite{Hor80,Sel92,Die02},\cite[Ch. 6]{HP10})
there is the following physical picture of the LP behind the above formal description.
In underlying systems, two different low-temperature ordered phases are possible,
and they are separated from the high-temperature disordered phase by a line \emph{($\ell$)} of
continuous second-order phase transitions.
One of them is homogeneous, here the non-zero averaged value of the order
parameter is the same throughout the system.
In another one, there is a periodic modulation of the order parameter along
certain anisotropy axes.
The line of transitions between these two ordered phases
terminates at the line \emph{($\ell$)}, and the point where they meet is just the LP.

Finally, let us note that an enormous amount of papers dealing
with different versions of "Lifshitz-like" theories appeared recently in
quantum field theory (QFT), particle physics, gravitation, and cosmology
(see \cite{AnsHal07,Ans08,Ans09,Hor09jhep,Hor09,Vis09,Calc09} and papers
that quote these references).
Here most of the work is concentrated on models in $d$-dimensional space-time
manifolds where higher space derivatives are allowed, but higher time derivatives are
forbidden and not generated by renormalization.
In the "Euclidean picture" of (\ref{EHI}), this corresponds to the situation
when one of the "perpendicular" components of $\bm x=(\bm x_{\perp},\bm x_\|)$
is associated with the time, while the remaining $d-m-1$ components of $\bm x_{\perp}$
as well as all $m$ "parallel" components are considered as spatial coordinates.
Obviously, the presence of anisotropy between time and space (as well as between
space coordinates alone) breaks the Lorentz symmetry in such models.
Hence the name "Lorentz violating (LV) theories" frequently used in the literature
\cite{AnsHal07,SVW09,FAlt11}.

Up to few exceptions \cite{AnsHal07,Ans08}, usually
one considers, as in \cite{Hor09,Vis09}, the case when
all space directions scale in the same manner.
In our notation this corresponds to fixing $d-m=1$ in (\ref{EHI}) and
working on the line $d=m+1$ of the $(m,d)$ plane.
On the other hand, of special interest in QFT are theories in the four-dimensional
space-time, which corresponds to the line $d=4$. The intersection point
of these two lines is $(3,4)$ where we have the usual "$d=3+1$" not forgetting
about the fourth-order derivatives along the three spatial directions.

Now, if we still are interested in integer values of $m$ by keeping $d=4$ but
allowing that some spatial directions scale like time, we land at $d=2+2$ and
$d=1+3$. Such possibilities have been considered in the classification of \cite{AnsHal07}.
Moreover, in the special case $m=1$, $d=4$ Anselmi \cite{Ans08} presents explicit
calculations and results for "Lifshitz type models" in the large-$n$ expansion,
which in the case of scalar fields completely parallel that of \cite{SPD05}
and of the present paper --- in the context of the condensed-matter physics
(for an explicit comparison of calculational techniques and results see Sec. \ref{d4}).

The model (\ref{EHI}) has several key features that attracted a growing interest
and led to numerous sophisticated generalizations in different fields of QFT.
First of all, the inclusion of fourth-order derivatives along $m$
spatial directions raises the upper critical dimension of $\phi^4$ models
$d^*$ from the usual $d^*=4$ at $m=0$ to $d^*(m)=4+m/2$, which can go up to
$d^*=8$ at $m=d$. This comes along with the fact that the renormalized values
of relevant Feynman diagrams are finite up to these relatively high dimensions.
The same can be seen as an improvement of ultraviolet convergence of Feynman integrals
(saying nothing about technical complications!)
in fixed dimensions with respect to their counterparts in conventional CP-like field theories.
For example, the same diagrams that usually diverge when $d>4$ are finite up to $d=5.5$
when $m=3$. A similar thing happens also to the
lower critical dimension $d_\ell(m)$ of models with continuous $O(n)$ symmetry.
As we discussed above, now we have $2\le d_\ell(m)=2+m/2\le4$.

Further, the model (\ref{EHI}) can be extended in two obvious ways.
It is possible to add terms containing two fields and derivatives higher than four
in the parallel subspace.\footnote{Early work in this direction
\cite{NCS76,NTCS76,Sel77,FH93} in condensed-matter physics did not receive much attention.}
Demanding that gradient terms of highest order,
say $2z$,\footnote{Different values of $z$ lead to different classical values
of the anisotropy exponent $\theta_0=1/z$. However, as we already discussed above,
the critical exponent $\theta$ differs from its classical value $\theta_0$ due to the
non-trivial renormalization of the interacting theory. Hence, we find it misleading
to identify the derivative's power in the Hamiltonian with the
critical exponent as it is sometimes done in the literature.}
scale in the same way as $(\nabla_{\perp}\bm{\phi})^2$
yields the modified lines of critical dimensions,
$d^*_z(m)=4+m(z-1)/z$ and $d_{\ell,z}(m)=2+m(z-1)/z$.
Moreover, higher powers of fields without gradients can be added, and
models with $g_s\phi^{s}$ interaction terms can be considered
apart from the usual $\phi^4$ one.
The engineering $\mu$ dimension of the effective expansion parameter
$\bar g_s$ is $[\bar g_s]_\mu=\left(d^*_{z,s}(m)-d\right)(s-2)/2$ where
$d^*_{z,s}(m)=2s/(s-2)+m(z-1)/z$.

Clearly, such generalizations allow to consider a very large amount of theories
with different $d$, $m$, $z$, and $s$, which will have very different
properties depending on chosen set of these parameters.
In particular, we see that the "weighted" power counting raises the
engineering dimensions of coupling constants at $\phi^s$ when $m>0$ and $z>1$.
Hence it is possible that certain non-renormalizable Lorentz-symmetric models
can become renormalizable at the same $d$ when the Lorentz symmetry is broken.
A classification of LV theories
involving scalar and fermion fields has been undertaken in \cite{AnsHal07}
and generalized to gauge fields in \cite{Ans09a,Ans09b}.

To give some examples we mention a scalar $\phi^{10}$ model with $z=2$ and $m=3$
whose upper critical dimension $d^*$ is $d^*_{2,10}(3)=4$, see \cite[Eq. (2.14)]{AnsHal07},
\cite[Eqs. (2.4)-(2.5)]{IRS09}.
Another example with $d^*=4$ is a $\phi^6$ model with $z=2$ and $m=2$
\cite[Eq. (2.15)]{AnsHal07}. Reference \cite{IRS09} gives explicit one-loop
calculations for a similar $\phi^6$ model with $z=2$, $m=4$ and $d^*=5$.
The last two instances belong to the class of theories describing the
$m$-axial tricritical LPs in strongly anisotropic systems with
short-range interactions. These have been considered some time ago in
\cite{Deng85,ADH85,ADH87}. Further generalization involving extra
long-ranged uniaxial dipole-dipole interactions can be found in \cite{FM93,AF96}.

Similarly as we noted it above by discussing the LP at $z=2$,
the field $\bm\phi$ becomes dimensionless at $d=d_{\ell,z}(m)$.
In this case $[\bar g_s]_\mu=2$ for any $m$ and $z$ and independently of $s$.
Along with the constraint $d=m+1$, the condition
$d=d_{\ell,z}(m)$ is satisfied at $z=m$. In the four-dimensional
space-time $d=3+1$ this reduces to $z=3$.
Such kind of observations led Horava \cite{Hor09,Hor11,HorMT10}
(see also \cite{Vis09,SVW09,Vis11})
to his seminal formulation of the power-counting renormalizable
quantum field theory of "gravity at a Lifshitz point" in $3+1$ dimensions.
However, "beyond power-counting renormalizability... explicit calculations have so far
been extremely limited" \cite{Vis11}. A few examples that give some taste are
\cite{IRS09,AnsT10,NMT12}.

The aim of the present paper is an explicit large-$n$
calculation of critical exponents $\nu_{L2}$ and $\nu_{L4}$ that
control the behavior of correlation lengths $\xi_\perp$ and $\xi_\|$
at $m$-axial Lifshitz points.
These correlation lengths are related to directions perpendicular
and parallel to modulation axes of the periodically ordered low-temperature phase.
We have $\xi_\perp\sim|\tau|^{-\nu_{L2}}$ and $\xi_\|\sim|\tau|^{-\nu_{L4}}$
for small reduced temperature deviations $\tau\equiv(T-T_{LP})/T_{LP}$ from the LP.
The other exponents $\gamma_L$, $\alpha_L$, and $\beta_L$ derived below
give the thermal behavior of the susceptibility,
specific heat, and order parameter.
The present work extends that of  \cite{SPD05}
where the correlation
critical exponents $\eta_{L2}$ and $\eta_{L4}$ have been
calculated to first non-trivial order of the large-$n$ expansion.

Hopefully, the present rather technical  communication along with
\cite{DS00,SD01,DSZ03,SPD05,BDS10,RDS11} could be useful for
the bright community interested nowadays in Lifshitz-type theories.

Before going to detail of our new calculations we briefly summarize some main results
of \cite{SPD05}.

\section{Lifshitz point's correlation exponents at large $\bm n$}

Below, as well as in determining the exponents
$\eta_{L2}$ and $\eta_{L4}$ in \cite{SPD05}, we employ the method of \cite{VPH81a}.
It consists in deriving the
$1/n$ expansions from self-consistent equations for full correlation functions
of the interacting theory. The actual calculations are done in the
critical (massless) theory using the dimensional regularization,
see \cite[Ch. 4.35, 4.38]{Vas98}.
Directly at the LP, the corresponding theory is given by the LGW Hamiltonian
(cf. (\ref{EHI}))
\begin{equation}\label{EH}
\mathcal H_{LP}[\bm{\phi}]
=\frac{1}{2}\,\int d^{d-m}r\int d^mz\left[(\nabla_{\bm{r}}\bm{\phi})^2
+(\Delta_{\bm z}\bm{\phi})^2\right]
+\frac{\lambda}{8}\,{\int}d^dx\,(|\bm\phi|^2)^2\,.
\end{equation}
The values $\tau_0^{LP}$ and $\rho_0^{LP}$, which locate the
position of LP \cite{Die02,Die05,SPD05} and differ from zero beyond the
dimensional regularization, are ignored in the following calculation, similarly as
it was done in \cite{SPD05}.
It is assumed  that $\lambda\propto 1/n$ for $n\to\infty$ as in the usual $\phi^4$ model
\cite[Ch. 2.19]{Vas98}. In the construction of the large-$n$ expansion, this
is not affected by the present modification of gradient terms.
For brevity, we use the notation $\bm r\equiv\bm x_{\perp}$, $\bm z\equiv\bm x_\|$
and hide the parameter $\sigma_0$ by appropriate
rescaling of the model (cf. \cite{Vis09}).
Implications of using the massless theory are discussed in Sec. \ref{APC}.
The applicability of the large-$n$ expansion to physically accessible
LPs is discussed in \cite{SPD05}.

Deviations from the LP can be effected in two ways.
First of all, the temperature variation will lead to the perturbation
of the Hamiltonian $\mathcal H_{LP}[\bm{\phi}]$ by the term
\be\label{O2}
\Delta\mathcal H_\tau[\bm{\phi}]=\frac{\tau}{2}\int d^dx\,|\bm\phi(x)|^2\,.
\ee
Its inclusion is needed in order to derive the thermal exponents
$\nu_{L2}$, $\nu_{L4}$, $\gamma_L$, and so on, which is the main objective
of the present paper.
On the other hand, in order to describe the crossover from the LP,
the term
\be\label{Or}
\Delta{\mathcal H}_\rho[\bm{\phi}]=\frac{\rho}{2}\int d^dx\,
(\nabla_{\bm{z}}\bm{\phi})^2
\ee
has to be included into the LGW Hamiltonian.
Taking it into account would open access to the crossover exponent $\varphi$
and the modulation wave-vector exponent $\beta_q$.
This is beyond of the scope of the present paper.

In the absence of perturbations, the long-wave asymptotic of the
full pair correlation function in the momentum representation
$\tilde G^{(\rm as)}_\phi(p,q)$ obeys the anisotropic scaling via
\be\label{Gold}
\tilde G^{(\rm as)}_\phi(p,q)=
p^{-2+\eta_{L2}}\tilde G^{(\rm as)}_\phi(1,q p^{-\theta})
=q^{-4+\eta_{L4}}\tilde G^{(\rm as)}_\phi(p q^{-1/\theta},1)\,.
\ee
Here only two of the three exponents are independent.
The anisotropy index $\theta$ is defined as
$\theta=(2-\eta_{L2})/(4-\eta_{L4})=\nu_{L4}/\nu_{L2}$.
In the large-$n$ limit, the non-trivial correlation function
$\tilde G^{(\rm as)}_\phi(p,q)$
reduces to the free-theory propagator $\tilde{G}^{(0)}_\phi(p,q)$:
\be\label{Sphl}
\lim_{n\to \infty}\tilde G^{(\rm as)}_\phi(p,q)=\left(p^2+q^4\right)^{-1}\,.
\ee
Its explicit form can be read off from the first term of (\ref{EH})
after a Fourier transformation.

The main result of \cite{SPD05} have been the large-$n$ expansions
\be\label{Mainr}
\eta_{L2}={\eta_{L2}^{(1)}\over n}+O\big(n^{-2}\big),\;
\eta_{L4}={\eta_{L4}^{(1)}\over n}+O\big(n^{-2}\big),\;
\theta=\frac{1}{2}+{\theta^{(1)}\over n}+O\big(n^{-2}\big)
\ee
obtained for generic number of anisotropy axes $0\le m\le d$
in $d$-dimensional space. Their $1/n$ coefficients have been
derived by considering the self-consistent
equation for the function $\tilde G^{(\rm as)}_\phi(p,q)$.
In a graphical form, it may be represented as
\be\label{nge}
\left[\tilde G^{(\rm as)}_\phi(p,q)\right]^{-1}=\frac{2}{n}\;
\raisebox{-11pt}{\includegraphics[width=39pt]{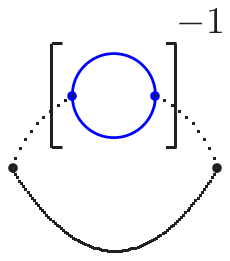}}
\ee
where solid lines denote the full propagators
$\tilde G^{(\rm as)}_\phi(p,q)$. This is a symbolic picture
of the self-consistent equation
written up explicitly in (23)-(24) of \cite{SPD05}.
Solving this equation for $q=0$ and for $p=0$ yielded
the $1/n$ expansion coefficients
\bea\label{E2c}
&&\eta_{L2}^{(1)}=2{K_{d-m}\over d-m}
\int_{\bm{q}}^{(m)}{4-(d-m)(1+q^4)\over (1+q^4)^3}\;{1\over I(1,q)},
\\\label{E41}
&&\eta_{L4}^{(1)}={2K_m\over m(m+2)}\,
\int_{\bm{p}}^{(d-m)}{{\mathcal P}_2(p^2)\over (p^2+1)^5}\;{1\over I(p,1)}
\qquad\mbox{with}
\eea
$$
\mathcal P_2(p^2)=3(8-m)(6-m)+5(m^2+2m-96)p^2+(m^2+50m+144)p^4-m(m+2)p^6,
$$
and
\be\label{Thc}
\theta^{(1)}=-\eta_{L2}^{(1)}/4+\eta_{L4}^{(1)}/8\,.
\ee
The function $I(p,q)$,
\begin{equation}\label{Ipq}
I(p,q)=\int_{\bm{p}'}^{(d-m)}\int_{\bm{q}'}^{(m)}
\frac{1}{{p^\prime}^2+{q^\prime}^4}\;
\frac{1}{|\bm{p}'+\bm{p}|^2+|\bm{q}'+\bm{q}|^4}\;,
\end{equation}
represents the "elementary bubble" \cite{Ma73} at LP.
It associates with the Feynman diagram
\raisebox{-3pt}{\includegraphics[width=30pt]{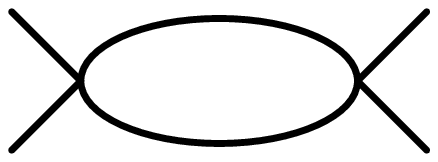}}\,
of the $\phi^4$ theory.
For any $D$-dimensional momentum integrals we use the notation
\be\label{KN}
\int_{\bm k}^{(D)}f(\bm k)\equiv\int\frac{d^Dk}{(2\pi)^D}f(\bm k).
\ee
When the integrand is rotationally invariant, the angular integration
is trivial and we have
\be\label{Kid}
\int_{\bm{k}}^{(D)}f(k^2)=K_D\int_0^\infty d k\, k^{D-1}f(k^2),\quad\mbox{where}\quad
K_D\equiv\frac{S_D}{(2\pi)^D}=2\frac{(4\pi)^{-D/2}}{\Gamma(D/2)}
\ee
and $S_D=2\pi^{D/2}/\Gamma(D/2)$
is the surface area of a $D$-dimensional sphere of unit radius.

In \cite{SDP08} we have shown that the $1/n$ coefficients (\ref{E2c}) and
(\ref{E41}) agree for arbitrary $0\le m\le d$ with the $O(\ve^2)$
terms of the epsilon expansion of $\eta_{L2}$ and $\eta_{L4}$
obtained previously in \cite{DS00} and \cite{SD01}.
Let us recall that in the LP theory the upper critical dimension is given
by the line $d^*(m)=4+m/2$ and $\ve=d^*(m)-d$.

\section{Identification of the critical exponent $\bm{\nu_{L2}}$}\label{S1}

Up to now, we discussed the theory
(\ref{EH}) \emph{exactly} at the LP.
Further, we consider a small temperature deviation from criticality.
Following \cite{VPH81a}, we take into account
the term (\ref{O2}) by treating it as a small perturbation.
The classical momentum dimension of the variable $\tau$ is
$[\tau]_0=2$. Its full scaling dimension\footnote{For a complete
renormalization-group analysis of the critical behavior
at LP see \cite{DS00}.} is $1/\nu_{L2}$.
In the spherical limit \cite{HLS75n},
$1/\nu_{L2}^\infty=d-m/2-2$.
Our aim will be to calculate the $1/n$ correction to this value.

Let us consider the theory involving the "critical" LGW Hamiltonian
(\ref{EH}) and its perturbation (\ref{O2}).
In the presence of $\Delta\mathcal H_\tau$,
the full propagator $\tilde G^{(\rm as)}_\phi$ gets the new
temperature variable $\tau$. Now, it obeys anisotropic scaling via
\be\label{Srb}
\tilde G^{(\rm as)}_\phi(p,q;\tau)=b^{2\tilde\Delta_\phi}
\tilde G^{(\rm as)}_\phi(bp,b^\theta q;\tau b^{1/\nu_{L2}})
\quad\mbox{with}\quad 2\tilde\Delta_\phi=2-\eta_{L2}\,.
\ee
This implies that the generalized homogeneous function
$\tilde G^{(\rm as)}_\phi$ can be expressed in different
scaling representations. In particular, for $b=p^{-1}$,
\be\label{Sr}
\tilde G^{(\rm as)}_\phi(p,q;\tau)=p^{-2\tilde\Delta_\phi}
\tilde G^{(\rm as)}_\phi(1,q p^{-\theta};\tau p^{-1/\nu_{L2}})\,.
\ee
Alternatively, the powers $q^{-4+\eta_{L4}}$ (cf. (\ref{Gold}))
or $\tau^{-\gamma_L}$ can be scaled out, accompanied by
appropriate scaling functions.
Following the approach of \cite{VPH81a}, we have to consider
infinitesimal temperature deviations form the critical theory and
follow the changes they produce in the self-consistent equations.

As usual, it is expected that the asymptotic small-$\tau$ expansion
of $\tilde G^{(\rm as)}_\phi(p,q;\tau)$ at fixed finite momenta
$p$ and $q$ contains both integer and non-integer powers of $\tau$.
Including only the first few terms, this expansion can be written as
\bea\label{Corz}
&&
\tilde G^{(\rm as)}_\phi(p,q;\tau)\sim p^{-2\tilde\Delta_\phi}
\times\\&&\nonumber
\left[D_0(qp^{-\theta})+\tau p^{-1/\nu_{L2}}D_1(q p^{-\theta})
+\tau^{1-\alpha_L} p^{-(1-\alpha_L)/\nu_{L2}}\hat D(q p^{-\theta})
+O(\tau^2)\right].
\eea
The specific heat exponent $\alpha_L$ is \cite{HLS75n}
\be\label{Alinf}
\alpha_L={d-m/2-4\over d-m/2-2}+O(n^{-1})=-\frac{\ve}{2-\ve}
+O(n^{-1})<0,
\ee
and the power $1-\alpha_L=2/(2-\ve)+O(n^{-1})$ of $\tau$
in the second correction term is greater than $1$.
The powers of $p$ are given by
\be\label{Twopos}
\frac{1}{\nu_{L2}}=d-\frac{m}{2}-2+O(n^{-1})=2-\ve+O(n^{-1})\quad\mbox{and}\quad
\frac{1{-}\alpha_L}{\nu_{L2}}=2+O(n^{-1}),
\ee
while in the epsilon expansion both of them start with $2+O(\ve)$.
The scaling functions $D_0$, $D_1$, and $\hat D$ possess finite limits at $q=0$,
and in this case
\be\label{Corp}
\tilde G^{(\rm as)}_\phi(p,0;\tau)\sim p^{-2\tilde\Delta_\phi}
\left[A{+}B\tau p^{-1/\nu_{L2}}
{+}C\tau^{1-\alpha_L} p^{-(1-\alpha_L)/\nu_{L2}}{+}O(\tau^2)\right].
\ee
This is in full analogy with the well-known formulas from the CP theory
\cite{BAZ74,FA74,AA74,ZJ89,Vas98}.
There, the constant coefficients $A$, $B$, and $C$ have been calculated
both in the epsilon \cite{BAZ74,FA74} and large-$n$ \cite{AA74} expansions.
We shall not quote their explicit expressions, but only note that for $n$ large,
$B=O(1/n)$ while\footnote{In Eqs. (4.108) of \cite{Vas98} the
coefficients $C_2$ and $C_3$ are erroneously interchanged.}
$A$ and $C=const+O(1/n)$.
This is a quite general property
concerning only the $n$ dependencies in $\phi^4$ theories.
It certainly holds also in the LP theory.
This implies that in the large-$n$ limit (\ref{Corp}) reduces to
\be\label{Aes}
\lim_{n\to\infty}\tilde G^{(\rm as)}_\phi(p,0;\tau)\sim
p^{-2}\left[A_\infty+C_\infty\tau^{1-\alpha_L^\infty}
p^{-2}+O(\tau^{2(1-\alpha_L^\infty)})\right]\,.
\ee

On the other hand, $\tilde G^{(\rm as)}_\phi(p,q;\tau)$
at $n\to\infty$ gives the Gaussian propagator
\be\label{Ses}
\tilde{G}^\infty_\phi(p,q;r_\infty)=\left(p^2+q^4+r_\infty\right)^{-1},
\ee
where $r_\infty\sim\tau^{\gamma_L^\infty}$
is the inverse susceptibility of the spherical model with \cite{HLS75n}
$\gamma_L^\infty=2/(d-m/2-2)=2/(2-\ve)$. We see that
in the large-$n$ limit the susceptibility exponent $\gamma_L^\infty$
coincides with the power $1-\alpha_L^\infty$
as it follows from (\ref{Alinf}).
Therefore, the small-temperature expansion represented by (\ref{Aes})
directly matches an analogous expansion of the function
$\tilde{G}^\infty_\phi(p,q;r_\infty)$ from (\ref{Ses}). Indeed, hence we get
\be\label{Ges}
\tilde{G}^\infty_\phi(p,0;r_\infty)=
p^{-2}\left[1-r_\infty p^{-2}+O(r_\infty^2)\right]\,.
\ee
Since the correction terms in (\ref{Aes}) and (\ref{Ges}) are the same,
by comparing these equations we see that $A_\infty=1$, and
$-C_\infty$ is actually the amplitude of the inverse susceptibility
in the large-$n$ limit: $r_\infty=-C_\infty\tau^{\gamma_L^\infty}$.

A summary is that the correction term
$\sim\tau^{1-\alpha_L} p^{-(1-\alpha_L)/\nu_{L2}}$
in the asymptotic expansions (\ref{Corz}) and (\ref{Corp})
of the full propagator $\tilde G^{(\rm as)}_\phi$
directly matches in the large-$n$ limit the linear $r_\infty p^{-2}$
contribution of (\ref{Ges}) stemming from the spherical model's
correlation function $\tilde G^\infty_\phi(p,q)$.
By contrast, the linear $O(\tau)$ correction of (\ref{Corp}) disappears at
$n\to\infty$ due to vanishing of its amplitude, and it has no counterpart
in the spherical limit.

From the above consideration we draw an important practical consequence:
In the following we shall parametrize the function
$\tilde G^{(\rm as)}_\phi$ by the temperature variable $\tau^{1-\alpha_L}$
instead of $\tau$. Thus we shall write
\be\label{nsa}
\tilde G^{(\rm as)}_\phi(p,q;\tau)\equiv\hat G^{(\rm as)}_\phi(p,q;\hat\tau)
=p^{-2\tilde\Delta_\phi}
\hat G^{(\rm as)}_\phi(1,q p^{-\theta};\hat\tau p^{-2\zeta})
\ee
where
\be\label{Choice}
\hat\tau\equiv\tau^{1-\alpha_L}\quad\quad\mbox{and}\quad\quad
2\zeta\equiv\frac{1-\alpha_L}{\nu_{L2}}\,.
\ee
Directly at the LP we have the obvious relation
$\hat G^{(\rm as)}_\phi(p,q;0)=\tilde G^{(\rm as)}_\phi(p,q;0)
\equiv\tilde G^{(\rm as)}_\phi(p,q)$.

Assume (cf. (\ref{Twopos})) that the exponent $2\zeta$ has a large-$n$ expansion
\be\label{Zetadef}
2\zeta=2\zeta_0+{\zeta_1\over n}+O(n^{-2})\quad\mbox{with}\quad \zeta_0=1
\ee
and unknown $1/n$ coefficient $\zeta_1$. Its determination
will be the goal of the following section.
To get a direct relation between $\zeta_1$ and the correlation length
exponent $\nu_{L2}$ we use the hyperscaling law \cite{HLS75}
\be\label{Hyper}
\alpha_L=2-(d-m)\nu_{L2}-m\nu_{L4}=2-(d-m+\theta m)\nu_{L2}
\ee
in (\ref{Choice}). This yields
\be\label{n2def}
{1\over\nu_{L2}}=d-\frac{m}{2}-2+m\,{\theta^{(1)}\over n}-
{\zeta_1\over n}+O(n^{-2})\,.
\ee

We shall need also relations providing the one-to-one correspondence
between the full propagator and its spherical limit's counterpart.
At $n\to\infty$, the function $\hat D(y)$
that appeared for the first time in (\ref{Corz}), is given by
\be\label{Sphd}
\lim_{n\to\infty}\hat D(y)=\frac{C_\infty}{(1+y^4)^2}\,,
\ee
which follows from (\ref{Ses})--(\ref{Ges}).
The constants $C$ and $A$ from (\ref{Corp}) are defined as
\be\label{Coa}
C=\hat D(0)\qquad\mbox{and}\qquad A=D_0(0)=\hat G^{(\rm as)}_\phi(1,0;0).
\ee

In the following section we shall find the value of $\zeta_1$, and hence
the critical exponent $\nu_{L2}$ up to order $O(1/n)$.

\section{Self-consistent equation and critical exponents}\label{S2}

In the presence of a perturbation,
the self-consistent equations for full correlation functions
are of the same general form as that in the theory at criticality
\cite{VPH81a,Vas98}.
That is, when the deviation $\Delta\mathcal H_\tau [\bm{\phi}]$ from the LP
is included, the self-consistent equation to be solved below,
is again given by (\ref{nge}).
The only essential difference is the presence of an additional
parameter $\hat\tau$ as the new argument of the propagator
$\hat G_\phi^{({\rm as})}$. Explicitly we have
\begin{equation}\label{Gphias}
\big[\hat G^{(\rm as)}_\phi(p,q;\hat\tau)\big]^{-1}=
\frac{2}{n}\int_{\bm{p}'}^{(d-m)}\int_{\bm{q}'}^{(m)}
\frac{\hat G^{(\rm as)}_\phi(|\bm{p}'+\bm{p}|,
|\bm{q}'+\bm{q}|;\hat\tau)}{F(p',q';\hat\tau)}
\end{equation}
with
\begin{equation}\label{BubbleI}
F(p,q;\hat\tau)=\int_{\bm{p}'}^{(d-m)}\int_{\bm{q}'}^{(m)}
\hat G^{(\rm as)}_\phi(|\bm{p}'+\bm{p}|,
|\bm{q}'+\bm{q}|;\hat\tau) \,\hat G^{(\rm as)}_\phi(p^\prime,q^\prime ;\hat\tau)
\end{equation}
where we use the representation (\ref{nsa}) for the full correlation function
$G_\phi^{(\rm as)}$.

Following the lines of \cite{VPH81a}, we expand the both sides of (\ref{Gphias})
to linear order in $\hat\tau$. Equating the zeroth-order terms reproduces
the self-consistent equation (23) of \cite{SPD05}, which yields
$\eta_{L2}$ and $\eta_{L4}$ to order $O(1/n)$.
Matching the $O(\hat\tau)$ contributions leads to the equation
\be\label{E12}
\partial_{\hat\tau}\hat G^{(\rm as)-1}_\phi
\left.(p,q;\hat\tau)\right|_{\hat\tau=0}
=\frac{2}{n}\left[E_1(p,q)-2E_2(p,q)\right]
\ee
with
\be\label{E1}
E_1(p,q)=\int_{\bm{p}'}^{(d-m)}\int_{\bm{q}'}^{(m)}
\frac{\partial_{\hat\tau}\hat G^{\rm (as)}_\phi(|\bm{p}'+\bm{p}|,
|\bm{q}'+\bm{q}|;\hat\tau)|_{\hat\tau=0}}{F(p',q';0)}
\ee
and
\be\label{E2}
E_2(p,q)=\int_{\bm{p}'}^{(d-m)}\int_{\bm{q}'}^{(m)}
\hat G^{(\rm as)}_\phi(|\bm{p}'+\bm{p}|,|\bm{q}'+\bm{q}|)\;
\frac{W(p',q')}{[F(p',q';0)]^2}\,.
\ee
The function $W$ appearing in the integrand of $E_2(p,q)$ is given by
\begin{equation}\label{Wdef}
W(p,q)=\int_{\bm{p}'}^{(d-m)}\int_{\bm{q}'}^{(m)}
\hat G^{(\rm as)}_\phi\left.(|\bm{p}'+\bm{p}|,|\bm{q}'+\bm{q}|)\;
\partial_{\hat\tau}\hat G^{\rm(as)}_\phi
(p^\prime,q^\prime;\hat\tau)\right|_{\hat\tau=0}.
\end{equation}
In (\ref{E1}) and (\ref{Wdef}), $\partial_{\hat\tau}$
means the partial derivative with respect to $\hat\tau$.

All functions appearing in above equations are generalized homogeneous
functions. They can be written in the scaling representations similar
to that of (\ref{Gold}). Thus, using (\ref{Gold}) and (\ref{nsa}) we write
$W(p,q)$ as
\be\label{Scw}
W(p,q)=p^{d-m+\theta m-4\tilde\Delta_\phi-2\zeta}\,W(1,p^{-\theta}q)\,.
\ee
This will be needed below, along with scaling
representation (\ref{Gold}) for $\tilde G^{(\rm as)}_\phi(p,q)$ and \cite{SPD05}
\be\label{Scf}
F(p,q)\equiv F(p,q;0)=p^{d-m+\theta m-4\tilde\Delta_\phi}\,
F(1,p^{-\theta}q)\,.
\ee

Similarly as in \cite{SPD05}, we consider the equation (\ref{E12})
at zero external momentum $q=0$.
Then, taking into account (\ref{Corp}), we obtain
$-p^{2\tilde\Delta_\phi-2\zeta}C/A^2$ on the left.
The constants $C$ and $A$ are given by (\ref{Coa}).
Using the scaling representations for all functions involved in $E_1(p,0)$
and $E_2(p,0)$, we scale out the same power $p^{2\tilde\Delta_\phi-2\zeta}$
on the right.
Matching the amplitudes at $p^{2\tilde\Delta_\phi-2\zeta}$ on both sides
we get the equation
\be\label{Me}
-\frac{C}{A^2}=\frac{2}{n}\left[E_1(1,0)-2E_2(1,0)\right]\,.
\ee
Here we have
\bea\label{E11}
&&E_1(1,0)=\int^{(d-m)}_{\bm{p}}\,
\frac{p^{4\tilde\Delta_\phi-(d-m)}}{|\bm{p}
+\bm{1}|^{2\tilde\Delta_\phi+2\zeta}}\,
\int^{(m)}_{\bm{q}}\frac{\hat D\big(p^\theta|\bm{p}+\bm{1}|^{-\theta}\,q\big)}
{F(1,q)}\,,
\\\label{E21}
&&E_2(1,0)=\int^{(d-m)}_{\bm{p}}\,
\frac{p^{4\tilde\Delta_\phi-2\zeta-(d-m)}}{|\bm{p}
+\bm{1}|^{2\tilde\Delta_\phi}} \,
\int^{(m)}_{\bm{q}}
\hat G^{(\rm as)}_\phi\Big(1,\frac{p^\theta q}{|\bm{p}+\bm{1}|^\theta}\Big)
\frac{W(1,q)}{[F(1,q)]^2}.
\eea

As discussed in the preceding section, at large $n$ the amplitudes
$A$ and $C$ are of the same, zeroth order in $1/n$, and
the whole equation (\ref{Me}) has to be of order $O(1)$. Hence,
the integrals (\ref{E11}) and (\ref{E21}) must have simple poles in $1/n$
for the compatibility of the left- and right-hand sides
of the matching condition (\ref{Me}).
The same property has also the analogous CP's
self-consistent equation \cite{VPH81a}\footnote{By contrast to the
present work, the calculations of \cite{VPH81a}
have been done in the coordinate space.
That is why, there is no direct one-to-one correspondence in
intermediate details of both calculations, even in the isotropic limit $m=0$.}
derived in the momentum representation.

Similarly as in the calculation of $\eta_{L2}$ in \cite{SPD05},
the origin of the singularities of momentum integrals in
$E_1(1,0)$ and $E_2(1,0)$ at $n\to\infty$ is the behavior
of their integrands at large momenta $p$.
Indeed, as $p\to\infty$, the arguments $p^\theta|\bm p+\bm 1|^{-\theta}q$
of the scaling functions $\hat D$ and $\hat G^{(\rm as)}_\phi$
in (\ref{E11}) and (\ref{E21}) tend to their limiting
value $q$, and thus the inner $q$ integrals are $p$-independent in this limit.
At the same time, the decoupled outer $d-m$-dimensional $p$ integrations involve
the powers $p^{2\tilde\Delta_\phi-2\zeta-(d-m)}$ both in (\ref{E11}) and (\ref{E21}).
They develop the ultraviolet pole singularities controlled by the small value
$(\eta_{L2}^{(1)}+\zeta_1)/n$ for large $n$.
This is easily seen by taking into account the
definitions of $\tilde\Delta_\phi$ and $\zeta$
from (\ref{Srb}) and (\ref{Zetadef}).
The upshot is that at large $n$ the matching condition (\ref{Me}) reduces to
\be\label{Rme}
-C_\infty=2 \mbox{Res}_{1/n\to 0}\left[E_1(1,0)-2E_2(1,0)\right]
\ee
where we used (see (\ref{Ges})) the large-$n$ limit $A_\infty=1$ on the left.

Of crucial importance here is the mutual cancellation of the $O(1)$ contributions
from $2\tilde\Delta_\phi$ and $2\zeta$, such that their difference is $O(1/n)$.
This is provided by the identification (\ref{Choice}) of $2\zeta$ along with
treating the full propagator $G^{(\rm as)}_\phi$ in the form (\ref{nsa}).
An attempt to use instead the more standard parametrization (\ref{Sr})
for $G^{(\rm as)}_\phi$ and to identify $2\zeta$ with $1/\nu_{L2}$
(see (\ref{Twopos})) would prevent the
appearance of $1/n$ poles in the integrals $E_1(1,0)$ and $E_2(1,0)$
due to the lack of this cancellation.
This is different from the situation in the CP theory.
There \cite{VPH81a}, \cite[Ch. 4.38]{Vas98}, the equation
for the "correction exponent" $2\lambda$ is symmetric under the changes
$2\lambda\leftrightarrow d-2\lambda$.
This implies that the scale dimension $2\lambda$
could be identified with both powers of correction contributions,
$1/\nu$ or $(1-\alpha)/\nu$. Indeed, owing to the hyperscaling law
$\alpha=2-d\nu$, one has $(1-\alpha)/\nu=d-1/\nu$.

As it was readily mentioned, the
residua of the integrals $E_1(1,0)$ and $E_2(1,0)$ at $n\to\infty$
depend on the sum of $1/n$ coefficients $\eta_{L2}^{(1)}$ and $\zeta_1$. We have
\be\label{E44}
E_1(1,0)={n K _{d-m}\over\eta_{L2}^{(1)}+\zeta_1}
\int^{(m)}_{\bm{q}}\left.\frac{\hat D(q)}{F(1,q)}\right|_{n\to\infty}+O(1)\,,
\ee
\be\label{E45}
E_2(1,0)={n K _{d- m}\over\eta_{L2}^{(1)}+\zeta_1}
\int^{(m)}_{\bm{q}}\left.\hat G^{(\rm as)}_\phi(1,q)\;
\frac{W(1,q)}{[F(1,q)]^2}\right|_{n\to\infty}+O(1)\,.
\ee
The value $\eta_{L2}^{(1)}$ has been calculated in \cite{SPD05} and quoted in
(\ref{E2c}). The new coefficient $\zeta_1$ can be determined via
(\ref{Rme}) using the pole parts of $E_1(1,0)$ and $E_2(1,0)$
from (\ref{E44}) and (\ref{E45}).
Reducing there the scaling functions in the integrands to their
spherical model's counterparts
via (\ref{Ses}) and (\ref{Sphd}) we obtain
\be\label{Epz}
\zeta_1=-\eta_{L2}^{(1)}-2K_{d-m}\int^{(m)}_{\bm q}\frac{1}{(1{+}q^4)^2}
\frac{1}{I(1,q)}
+4K_{d-m}\int^{(m)}_{\bm q}{1\over 1{+}q^4}{J(1,q)\over [I(1,q)]^2}.
\ee
The function $I(1,q)=\lim_{n\to\infty}F(p,q)$, has already been
encountered in (\ref{Ipq}). Similarly, $J(1,q)$ is defined by
\be\label{J1q}
J(p,q)=C_\infty^{-1}\lim_{n\to\infty}W(p,q)
=\int_{\bm{p}'}^{(d-m)}\int_{\bm{q}'}^{(m)}
{1\over(p'^2+q'^4)^2}{1\over|\bm p'+\bm p|^2+|\bm{q}'+\bm q|^4}.
\ee
The non-universal amplitude $C_\infty$, related to the normalization of the
temperature variable $\hat\tau$, cancels in the equation for $\zeta_1$ as
it should.

Let us introduce the short-hand notations for frequently appearing integrals:
\bea\label{nn1}
&&{\mathcal N}_k\equiv K_{d-m}\int_{\bm q}^{(m)}
{1\over(1+q^4)^k}{1\over I(1,q)}\,,
\quad k  \quad\mbox{integer,}\quad\ge 2\,;
\\&&\label{nn2}
{\mathcal N}_J\equiv K_{d-m}\int_{\bm q}^{(m)}{1\over 1+q^4}\;
{J(1,q)\over[I(1,q)]^2}\,.
\eea
In terms of these integrals we can write
\be\label{Z1fin}
\zeta_1=-\eta_{L2}^{(1)}-2{\mathcal N}_2+4{\mathcal N}_J
=-\frac{8}{d-m}{\mathcal N}_3+4{\mathcal N}_J
\ee
where the second equality follows by
eliminating the coefficient $\eta_{L2}^{(1)}$ via (\ref{E2c}).
Using (\ref{n2def}), we end up with
\be\label{nu2fin}
\nu_{L2}^{-1}=d-\frac{m}{2}-2+\frac{1}{n}\left[\eta_{L2}^{(1)}+
m\theta^{(1)}+2{\mathcal N}_2-4{\mathcal N}_J\right]+O(n^{-2}).
\ee
The $1/n$ coefficients $\eta_{L2}^{(1)}$ and $\theta^{(1)}$ are
known from \cite {SPD05} and quoted in (\ref{E2c}) and (\ref{Thc}).

This is the central result of the present paper.
The knowledge of $\nu_{L2}$ along with
$\eta_{L2}$ and $\eta_{L4}$, allows us to
derive critical exponents of the susceptibility
and parallel correlation length,
$\gamma_{L}$ and $\nu_{L4}$ through the scaling relations \cite{HLS75}
\be\label{Dsc}
\gamma_L=\nu_{L2}(2-\eta_{L2})=\nu_{L4}(4-\eta_{L4}).
\ee
From the first equality we get
\be\label{Gam}
\gamma_L={2\over d-m/2-2}\left(1-{2\over d-m/2-2}\;
C_\gamma\;\frac{1}{n}\right) +O(n^{-2}),
\ee
where
\be\label{Gamc}
C_\gamma=\frac{d-m}{4}\,\eta_{L2}^{(1)}+{m\over 16}\,\eta_{L4}^{(1)}
+{\mathcal N}_2-2{\mathcal N}_J
\ee
is the $1/n$ coefficient of the inverse value
$\gamma_L^{-1}{=}(d{-}m/2{-}2)/2+C_\gamma/n+O(n^{-2})$.
The critical exponent $\nu_{L4}$ is obtained from the
second equality of (\ref{Dsc}):
\be\label{Fn4}
\nu_{L4}=\frac{1}{2}\,{1\over d-m/2-2}\left(1-{2\over d-m/2-2}\;
C_{\nu_4}\;\frac{1}{n}\right) +O(n^{-2})
\ee
with
\be\label{Cn4}
C_{\nu_4}=\frac{d-m}{4}\,\eta_{L2}^{(1)}-
\frac{1}{8}(d-m-2)\eta_{L4}^{(1)}+{\mathcal N}_2-2{\mathcal N}_J\,.
\ee
Analogous results for $\alpha_L$ and $\beta_L$
follow from (\ref{Hyper}) and \cite{HLS75} $\beta_L=(2-\gamma_L-\alpha_L)/2$.

Below, our findings will be checked in two limits
$m\to 0$ and $m\to d$, and in $\ve$ expansion to order $O(\ve^2)$.
In the special case $m=1$ we shall give analytical calculations at
$d=4$ and numerical results at $d=3$.

\section{Large-$n$ expansions in isotropic limits}

In this section we consider the limits $m\to 0$ and $m\to d$.
In both cases the system loses its spatial anisotropy.
That's why we call them isotropic.

\subsection{Critical-point limit}\label{CPL}

When $m\to 0$, our system reduces to that with
isotropic short-range interactions in the vicinity of the usual CP.
In the same manner as in \cite{SPD05}, we obtain
from (\ref{nu2fin})
\be
\lim_{m\to 0}\nu_{L2}^{-1}=d-2+\eta+2K_d\lim_{m\to 0}\left[\frac{1}{I(1,0)}
-2\frac{J(1,0)}{I^2(1,0)}\right]\frac{1}{n}+O(n^{-2})\,.
\ee
Here $\eta=\eta^{(1)}/n+O(n^{-2})$ is the Fisher correlation exponent
at the CP. In writing it we used the result
$\lim_{m\to 0}\eta_{L2}^{(1)}=\eta^{(1)}$ of \cite{SPD05}.

The limit $m\to 0$ reduces $I(1,0)$ and $J(1,0)$ to standard integrals of the form
\begin{equation}\label{Vdef}
\int_{\bm k}^{(D)}\,
k^{-2a}\,|\bm{k}+\bm{1}|^{-2b}\equiv V_D(a,b)\,.
\end{equation}
In the dimensional regularization,
\be\label{Vexp}
V_D(a,b)=(4\pi)^{-D/2}\;\frac{\Gamma\left(D/2-a\right)}{\Gamma(a)}\;
\frac{\Gamma\left(D/2-b\right)}{\Gamma(b)}\;
\frac{\Gamma\left(a+b-D/2\right)}{\Gamma(D-a-b)}.
\ee
Thus, $I(1,0)$ and $J(1,0)$ are identified as $V_d(1,1)$ and $V_d(2,1)$,
and we get
\be\label{NCP}
\lim_{m\to 0}\nu_{L2}^{-1}=\nu^{-1}=d-2+2\,\frac{(d-1)(d-2)}{4-d}\;\eta^{(1)}
\frac{1}{n}+O(n^{-2})\,.
\ee
This coincides with the well-known large-$n$ expansion result for $\nu$,
the usual CP correlation length exponent: see e.g. \cite{Ma73} or
\cite[(4.294)--(4.295)]{Vas98}.

Along the same lines, the CP susceptibility exponent $\gamma$ is reproduced,
\bea\label{GMA}\nonumber
\lim_{m\to 0}\gamma_L=\gamma&=&\frac{2}{d{-}2}\left\{1-\frac{2}{d{-}2}\left[
\frac{d}{4}\eta^{(1)}{+}\frac{K_d}{V_d(1,1)}-2K_d\frac{V_d(2,1)}{V_d^2(1,1)}
\right]\frac{1}{n}\right\}{+}O(n^{-2})
\\&=&
\frac{2}{d-2}\left[1-\frac{3}{2}\frac{d}{4-d}\;\eta^{(1)}\frac{1}{n}
\right]+O(n^{-2})\,,
\eea
in full agreement with \cite{Ma73}.

\subsection{Isotropic LP limit}

Another non-trivial check is provided by the isotropic
LP (ILP) limit, $m\to d$. This would be useless for
$\eta_{L2}$ and $\nu_{L2}$ separately, since each of them
loses its physical significance at ILP.
But the exponent $\gamma_L$, while being
combined from $\eta_{L2}$ and $\nu_{L2}$ via (\ref{Dsc}),
is well defined and meaningful at $m=d$.
Its $m\to d$ limit is expected to reproduce the known result of \cite{HLS75n}
(see also \cite{IHB90}) derived in the ILP theory,
\be\label{Ghls}
\gamma_{\rm ILP}=\frac{4}{d-4}-\frac{1}{n}\frac{C_{HLS}}{(d/4-1)^2}\left[
1+\frac{(10{-}d)(d{-}5)}{3}+\frac{3(d{-}8)(d{-}6)}{4(d+2)}\right]
+O(n^{-2}).
\ee
The coefficient
\be\label{Chls}
C_{HLS}\equiv {\Gamma(d-4)\over\Gamma(d/2)\Gamma^2(d/2-2)
\Gamma(4-d/2)}
\ee
coincides with that in front of square brackets in Eq. (3) of \cite{HLS75n}.
Besides the simple explicit factors in (\ref{Gam}),
we must produce the non-trivial limit $m\to d$
of the combination $C_\gamma$ from (\ref{Gamc}).
This has to be compared with the coefficient
at $-(d/4-1)^{-2}/n$ in (\ref{Ghls}).

The first term in $C_\gamma$ disappears due to the factor $d-m$, while
$\eta_{L2}$ remains finite in the limit $m\to d$%
\footnote{The same holds for $\eta_{L2}$ in the $\ve$ expansion \cite{SD01}.}.
It is straightforward to deal also with the second term of (\ref{Gamc}),
since we have already worked out the limit $m\to d$ for $\eta_{L4}$ in
\cite{SPD05}. We have
\be\label{Con2}
\lim_{m\to d}{m\eta_{L4}^{(1)}\over 16}=\frac{3}{4} {(d-8)(d-6)\over d+2}C_{HLS}.
\ee
The explicit $d$-dependent factor here matches the
third term in square brackets of (\ref{Ghls}).

The third term in (\ref{Gamc}) is the integral ${\mathcal N}_2$ from (\ref{nn1}).
The function $I(1,q)$ in its integrand has the scaling property \cite{SPD05}
\be
I(1,q)=q^{-2\ve}I(q^{-2},1)
\ee
for generic $m$ and $d$, while $\ve=4+m/2-d$
(note that $\ve$ is not supposed to be small in this section).
With this in mind we write
\be
{\mathcal N}_2=K_{d-m}K_{m}\int_0^\infty dq\,
\frac{q^{-2(d-m)-1}}{(1+q^{-4})^2}\,{1\over I(q^{-2},1)}\,.
\ee
For vanishing $d-m$ and $q\to\infty$, the factor $q^{-2(d-m)-1}$ leads to
a logarithmic singularity of the last integral, while the remaining part
of its integrand remains regular. Hence
\be
\lim_{m\to d}{\mathcal N}_2=
\lim_{m\to d}K_{d-m}K_{m}{1\over 2(d-m)}{1\over I(0,1)}=
{K_{d}\over 2 V_d(2,2)}=C_{HLS}\,,
\ee
where the explicit value of $V_d(2,2)$ is given by (\ref{Vexp})%
\footnote{The same result can also be reached by considering
the limit $m\to d$ of the explicit expression
for $I(0,q)$ derived in Appendix B of \cite{SPD05} for generic $m$ and $d$.}.
Thus we reproduce the first contribution, $1$, in square
brackets of (\ref{Ghls}).

The treatment of the integral ${\mathcal N}_J$ is similar.
Here we have to take into account that
\be
J(1,q)=q^{-4-2\ve}J(q^{-2},1)
\ee
where the function $J(q^{-2},1)$ has a regular limit as $q\to\infty$, namely
$V_d(4,2)$. Hence,
\be
\lim_{m\to d}{-2\mathcal N}_J=-K_d{V_d(4,2)\over \left[V_d(2,2)\right]^2}=
\frac{1}{3}(10-d)(d-5)C_{HLS}
\ee
in agreement with the last remaining contribution of (\ref{Ghls}).

Thus, the large-$n$ expansion of the
susceptibility exponent $\gamma_L$ of the anisotropic
$m$-axial LP theory (\ref{Gam})--(\ref{Gamc}) possesses correct limits of the
fully isotropic LP and CP for generic values of $d$ in both cases.

From (\ref{Fn4})--(\ref{Cn4}) we also derive
the ILP limit of the parallel correlation length exponent $\nu_{L4}$:
$$
\nu_{\rm ILP}=\frac{1}{d-4}-\frac{1}{n}\frac{4 C_{HLS}}{(d-4)^2}\left[
1+\frac{(10-d)(d-5)}{3}+3\,\frac{(d-8)(d-6)}{d(d+2)}\right]
+O(n^{-2})\,.
$$
The expression in square brackets differs from that of (\ref{Ghls})
only by the factor $d$ in place of $4$ in the last denominator.

In the following section we shall verify our results
without switching off the spatial anisotropy, by checking them against
the existing $\ve$ expansions for $m$-axial LPs.

\section{Compatibility of $\bm{1/n}$ and $\bm\ve$ expansions}\label{APC}

In this section we show the equivalence of $1/n$ and $\ve$ expansions for $\nu_{L2}$
in their common region of validity.
The compatibility of these two types of expansions for other LP exponents
then follows from scaling laws (see e.g. \cite{DS00}) by taking into account
that for correlation exponents $\eta_{L2}$ and $\eta_{L4}$ this property
has already been proved in \cite{SDP08}.
First we give a rather formal development dealing with generic integral representations
of involved functions valid for arbitrary $m\in[0,d]$.
In the following subsection we provide an example of an analytical
calculation at $m=2$ where all these functions are taken in their explicit forms.

\subsection{Generic $m$}
The critical exponent $\nu_{L2}^{-1}$ was obtained to $O(\ve^2)$ in \cite{SD01}
using dimensional regularization and minimal subtractions of $\ve$ poles
for $n\in[0,\infty[$ and $m\in[0,d]$. For large $n$ it reduces to
$$
\nu_{L2}^{-1}=2-\ve+\frac{6}{n} \ve
-{\ve^2\over 2n}\Bigg[{m-4\over 8-m}\,j_\phi(m)+{j_\sigma(m)\over 8\,(m+2)}
+28 J_u(m)\Bigg]\ve^2+O(\ve^3,n^{-2}).
$$
The precise expressions for the functions $j_\phi(m)$, $j_\sigma(m)$, and
$J_u(m)$ in terms of single integrals,
as well as their explicit values at $m=2$ are listed in Section 3.2 of \cite{SD01}.

The same result can be expressed in terms of correlation exponents as
\be\label{Etan}
\nu_{L2}^{-1}=2-\ve+6\frac{\ve}{n}+
\eta_{L2}+m\Big(\theta-\frac{1}{2}\Big)-14 J_u(m){\ve^2\over n}+O(\ve^3,n^{-2}),
\ee
which has the structure very similar to that of (\ref{nu2fin}).
The factor $(\theta-1/2)\sim\ve^2/n$
represents the non-classical, anomalous part
of the anisotropy index $\theta$. The term containing it, as well as
$\eta_{L2}$, have direct counterparts in (\ref{nu2fin}).
The contribution $J_u(m)$ stems in part from the non-trivial two-loop
renormalization of the $\phi^4$ coupling constant through the Feynman diagram
\raisebox{-7pt}{\includegraphics[width=30pt]{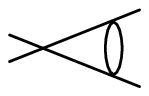}}
of the vertex function $\Gamma_4$. Another source of its appearance is the
similar Feynman graph (with glued left external lines)
of the two-point vertex function with a $\phi^2$
insertion, $\Gamma^{(2,1)}$. It is given by
\be\label{JUM}
J_u(m)=1-\frac{1}{2}C_E-\frac{1}{2}\psi\Big(2-\frac{m}{4}\Big)+j_u(m)
\ee
where $C_E=-\psi(1)\simeq 0.5772\ldots$ is the Euler's constant
and $\psi(z)$ is the digamma function,
the logarithmic derivative of the Gamma function (see e.g. \cite{AS,PBM1}).
The last term reads
\be\label{IJM}
j_u(m)=\frac{S_{4+m/2}}{F_{m,0}^2}\;b_m\int d^mu\,\Phi(u;m,0)\Theta(u;m)
\ee
where $b_m=2^{-4-m}\pi^{-3/2-m/4}$ as in (\ref{HPS}), the value $F_{m,0}$
can be easily read off from (\ref{Coni}), and
$S_D=(2\pi)^D K_D=2\pi^{D/2}/\Gamma(D/2)$ was introduced in (\ref{Kid}).
$\Phi(u;m,0)$ stands for the scaling function $\Phi$ from (\ref{Gx})
taken at $\ve=0$, and $\Theta(u;m)$ is defined in (\ref{Hx})--(\ref{TUM}).

Comparing equations (\ref{nu2fin}) and (\ref{Etan}) for $\nu_{L2}^{-1}$
we see that their equivalence for large $n$ and small $\ve$
requires that the $\ve$ expansion of
combination $2\mathcal N_2-4\mathcal N_J$ must be given by
\be\label{COMN}
2\mathcal N_2-4\mathcal N_J=6\ve-14 J_u(m)\ve^2+O(\ve^3)\,.
\ee
In the remaining part of this section we shall show that
this statement indeed holds true.

To deal with $\mathcal N_2$ and $\mathcal N_J$ defined in
(\ref{nn1})--(\ref{nn2}) we need the $\ve$ expansions of the functions
$I(1,q)$ and $J(1,q)$.
The integral $I(p,q)$ from ({\ref{Ipq}}) has an ultraviolet pole,
and its formal $\ve$ expansion can be written as
\be\label{Ie}
I(p,q)=p^{-\ve}\,\frac{c_{-1}}{\ve}\,\left[1+\ve\,i_0(q/\sqrt p)\right]+O(\ve)\,.
\ee
Using the $\ve$ expansion of $I(p,q)$ outlined in Appendix \ref{app1} we can write
\be\label{nDE}
\mathcal N_2=\ve\,\frac{K_{d-m}}{c_{-1}}\int_{\bm q}^{(m)}
{1\over(1+q^4)^2}\left[1-c_0\ve-\ve c_{-1}^{-1}I_0(1,q)+O(\ve^2)\right].
\ee
The part of $\mathcal N_2$ involving the first two constant terms from
square brackets reads
\be\label{nD1}
\mathcal N_2^{(1)}=\ve-\left[
1-\frac{1}{2}C_E-\frac{1}{2}\psi\Big(2-\frac{m}{4}\Big)\right]\ve^2+O(\ve^3).
\ee
Employing the integral representation (\ref{I1pq}) for $I_0(1,q)$
and interchanging the order of integrations
we obtain for the remaining part of $\mathcal N_2$
\be\label{nD2R}
\mathcal N_2^{(2)}=-\frac{K_{4-m/2}}{c_{-1}^2}\;\ve^2
\int d^dx [G^{(0)}(x)]^2\,\e^{-i\bm1\bm r}
\int_{\bm q}^{(m)}{\e^{-i\bm q\bm z}-1\over(1+q^4)^2}+O(\ve^3).
\ee
Here we recognize that the outer $q$
integral is the same as in the definition of
the function $H_0(r,z)$ introduced in (\ref{HR}). Hence, it can be viewed as
the Fourier back-transform of $H_0(r,z)$ in the subspace
of $r$ coordinates. This allows us to write
\be\label{nD2}
\mathcal N_2^{(2)}=-\frac{K_{4-m/2}}{c_{-1}^2}\;\ve^2
\int d^dx [G^{(0)}(x)]^2\,\e^{-i\bm1\bm r}
\int d^{d-m}r' H_0(r',z)\,\e^{i\bm1\bm r'}+O(\ve^3).
\ee
The integrals in (\ref{nD2}) can be safely evaluated at the upper critical
dimension $d^*=4+m/2$.
Here we use the scaling form (\ref{Gx}) for $G(x)$ along with (\ref{H1H})
for $H_0(r,z)$ and arrive at
\be\label{nD3}
b_m\,\int d^{4-\frac{m}{2}}r^{-4+\frac{m}{2}}\,\e^{-i\bm1\bm r}
\int d^{4-\frac{m}{2}}r'\,\e^{i\bm1\bm r'}
\int d^mu\,\Phi^2(u;m,0)\Theta\Big(u\sqrt\frac{r}{r'}\Big).
\ee
The structure of $r$ and $r'$ integrations
here is just the same as that encountered
in \cite{SDP08}, which produces a Dirac delta function
$(2\pi)^{4-m/2}\delta(\bm r-\bm r')$.
Thus we get for $\mathcal N_2^{(2)}$
$$
\mathcal N_2^{(2)}=-\frac{S_{4-m/2}}{c_{-1}^2}\;b_m\ve^2
\int d^mu \Phi^2(u;m,0)\Theta(u;m)+O(\ve^3)=-\ve^2\,j_u(m)+O(\ve^3)
$$
where in writing the last equality we took into
account the definition (\ref{IJM})
of the function $j_u(m)$ taken from \cite{SD01}.
Together with (\ref{nD1}) this gives
\be\label{N2T}
\mathcal N_2=\ve-J_u(m)\,\ve^2+O(\ve^3).
\ee

The treatment of the integral ${\mathcal N}_J$ requires additionally
an $\ve$ expansion of the function $J(p,q)$ from (\ref{J1q})
(see appendix \ref{Lex}).
At small $\ve$, this function
can be written in the scaling form analogous to (\ref{Ie}).
Again, both the pole and finite parts have to be taken into account
in the expansion
\be\label{Sj}
J(p,q)=\frac{h_{-1}}{\ve}\frac{p^{-2-\ve}}{1+q^4/p^2}
\left[1+\ve j_0(q/\sqrt p)\right]+O(\ve)\,.
\ee
By contrast to $I(p,q)$, the $1/\ve$ pole of $J(p,q)$ is
a function of $p$ and $q$ and has an infrared origin.
This can be seen from (\ref{J1q}) by power counting.
We cannot calculate neither $i_0(q/\sqrt p)$ nor $j_0(q/\sqrt p)$
analytically for arbitrary $m$. However, in the next section we shall
write down these functions explicitly for the special case $m=2$.

According to the decomposition (\ref{J3}) of the function $J(p,q)$ into
three parts, we have three contributions to $\mathcal N_J$.
The first of them, that involves $J^{(1)}(1,q)$ from (\ref{J1}), is
\be\label{NJR1}
\mathcal N_J^{(1)}=-K_{d-m}c_{-1}
\left[{1\over\ve}+\frac{C_E}{2}+\ln\sqrt\pi+O(\ve)\right]
\int_{\bm q}^{(m)}{1\over(1+q^4)^2}\frac{1}{I^2(1,q)}\,.
\ee
The integral in $\mathcal N_J^{(1)}$ differs from that in
$\mathcal N_2$ (see (\ref{nn1})) only by the power of $I(1,q)$
This implies that in the $\ve$ expansion, owing to the relation
\be
\frac{1}{I^2(1,q)}=\frac{2\ve}{c_{-1}}\,\frac{1}{I(1,q)}-\frac{\ve^2}{c_{-1}^2}
+O(\ve^3),
\ee
$\mathcal N_J^{(1)}$ can be simply expressed in terms of
$\mathcal N_2=\ve-\ve^2\,J_u(m)+O(\ve^3)$. This gives
\be\label{NJRO}
\mathcal N_J^{(1)}=\ve-2\mathcal N_2+\ve^2
\left[\frac{1}{2}\psi\Big(2-\frac{m}{4}\Big)
-\frac{C_E}{2}+\ln2\right]+O(\ve^3)\,.
\ee

The next contribution to $\mathcal N_J$, $\mathcal N_J^{(2)}$,
includes the $O(\ve^0)$ function $J^{(2)}$ from (\ref{JR2}).
Here only the leading $\ve^2/c_{-1}^2$ term from $1/I^2(1,q)$
is needed and we have
\be\label{NJR2}
\mathcal N_J^{(2)}=\ve^2\;\frac{K_{4-m/2}}{c_{-1}^2}
\int_{\bm q}^{(m)}{J^{(2)}(1,q)\over1+q^4}=\ve^2
\left[1-\psi\Big(2-\frac{m}{4}\Big)-\ln2\right]+O(\ve^3)\,.
\ee

In the last term, $\mathcal N_J^{(3)}$, we use the integral representation
(\ref{J3i}) for $J^{(3)}(1,q )$. Changing the order of
integrations, as in $\mathcal N_2^{(2)}$ (see (\ref{nD2R})), we obtain
\be\label{POI}
\mathcal N_J^{(3)}=\ve^2\;\frac{K_{4-m/2}}{c_{-1}^2}\;
b_m\int d^dx\,\Theta(u;m) G^{(0)}(x)\,\e^{-i\bm 1\bm r}
\int_{\bm q}^{(m)}{\e^{-i\bm q\bm z}\over1+q^4}+O(\ve^3)\,.
\ee
The Fourier transformations of (\ref{Gx}) imply that
the last $q$ integral can be written as
\be\label{DLR}
\int_{\bm q}^{(m)}{\e^{-i\bm q\bm z}\over1+q^4}=
\int d^{d-m}r'\, G^{(0)}(r',z)\e^{i\bm 1\bm r'}.
\ee
We use the scaling representation
(\ref{Gx}) for the free propagators $G^{(0)}$ in (\ref{POI}) and (\ref{DLR})
and change the integration
variable $\bm z$ via $\bm z=\bm u\sqrt r$ in the arguments of the scaling functions
$\Phi(z/\sqrt r)$ and $\Phi(z/\sqrt{r'})$.
Thus, up to a constant overall factor, the value of $\mathcal N_J^{(3)}$
is given by the following combination of integrals:
\be\label{NDT}
\int\!\! d^{4-\frac{m}{2}}r^{-2+\frac{m}{2}}\e^{-i\bm1\bm r}\!\!\!
\int\!\! d^{4-\frac{m}{2}}r'\,r'^{-2}\e^{i\bm1\bm r'}\!\!\!
\int\!\! d^mu\,\Theta(u;m)\Phi(u;m,0)\Phi\Big(u\sqrt\frac{r}{r'};m,0\Big).
\ee
Proceeding here by analogy with \cite{SDP08} and (\ref{nD3})--(\ref{N2T}) we get
\be\label{NJR3}
\mathcal N_J^{(3)}=\ve^2\,j_u(m)+O(\ve^3)\,.
\ee

Let us note that the results for both combinations of integrals in (\ref{nD3})
and (\ref{NDT}), which contain different powers of dummy variables
$r$ and $r'$, are the consequence of the following general theorem, which
can be proved along the lines of \cite{SDP08}:

If a $D$-dimensional Fourier transformation
$f(x)\leftrightarrow F(k)$ is defined via
\be
f(x)=\int_{\bm k}^{(D)}F(k)\e^{i\bm{kx}}\,,\mm{}\mm{}
F(k)=\int d^Dx f(x)\e^{-i\bm{kx}},
\ee
then
\be\label{GENT}
\int_{\bm s}^{(D)} s^{-\alpha} \e^{i\bm{vs}}
\int d^Dx x^{-\beta}g\Big(u\frac{x}{s}\Big)\e^{-i\bm{wx}}=
w^{-\alpha}v^{-\beta}g\Big(u\frac{v}{w}\Big)
\ee
provided that $\alpha+\beta=D$.

The integrals over $r$ and $r'$ in (\ref{nD3})
and (\ref{NDT}) are covered by (\ref{GENT}) with $v=w=1$ and
$\Theta(u\sqrt{r/r'})$ or $\Phi(u\sqrt{r/r'}):=g(u^2\,r/r')$, respectively.

Returning to $\mathcal N_J$, we collect terms given in (\ref{NJRO}),
(\ref{NJR2}), and (\ref{NJR3}), and obtain
\be\label{NJT}
\mathcal N_J=-\ve+3J_u(m)\,\ve^2+O(\ve^3)\,.
\ee
Finally, following (\ref{COMN}) we combine the values $\mathcal N_2$
and  $\mathcal N_J$ from (\ref{N2T}) and (\ref{NJT})
via $2\mathcal N_2-4\mathcal N_J$.
This really gives the combination
$6\ve-14 J_u(m)\ve^2+O(\ve^3)$ as required by (\ref{COMN})
for compatibility of $1/n$ and $\ve$ expansions for the critical
exponent $\nu_{L2}$.

In the following section we show how it works in a
closed-form calculation at $m=2$.

\subsection{Explicit calculation at $m=2$}\label{checkm2}

In this section we shall do analytical calculations for
$\nu_{L2}^{-1}$ in the case $m=2$ for $n\to\infty$ and $\ve\to 0$.
Contact will be made with the $\ve$ expansion result of \cite{SD01}.
Explicit expressions for relevant functions will be used
without appealing to formal Fourier transformations
employed in the preceding section.

For large $n$, the $O(\ve^2)$ expression \cite{SD01} for $\nu_{L2}^{-1}$
at $m=2$ reads
\be\label{R1n}
\nu_{L2}^{-1}(m=2)=2-\ve+6\frac{\ve}{n}+\left(\frac{4}{27}-
14\ln\frac{4}{3}\right){\ve^2\over n}+O(\ve^3,n^{-2})\,.
\ee
To see agreement of this formula with the result (\ref{nu2fin})
we shall explicitly expand it in $\ve$.

In the special case $m=2$ and $d^*(m=2)=5$, the scaling function
$\Phi(u)$ (see (\ref{Gx})) reduces to an ordinary Gaussian,
\begin{equation}\label{F2c}
\Phi(u;2,0)=(4 \pi)^{-2}\;\e^{-u^2/4}\,.
\end{equation}
Then, it is straightforward to
calculate $I_0(1,q)$ from (\ref{I1pq}) by using
\be \int d^2 u \; \Phi^2(u;2,0)\left(\e^{i\bm {q}\bm u\sqrt{r}}-1\right)=
\frac{1}{2}{1\over (4\pi)^3}\left(\e^{-\frac{1}{2} q^2 r}-1\right)\,.
\ee
Completing the remaining three-dimensional $r$ integral in (\ref{I1pq})
and combining its result with the constant $O(\ve^0)$ term from (\ref{Coni})
via (\ref{IFO}) we obtain
\be\label{F1q}
i_0(q)=\frac{1}{2}\Big[-q^2\,\arctan{2\over q^2}
-\ln{\Big(1+{q^4\over 4}\Big)}+2-C_E+\ln{4\pi}\Big].
\ee
This function has to be used in the finite part of the $\ve$ expansion of
$I(1,q)$ given by (\ref{Ie}) at $m=2$. In this case $c_{-1}=1/(32\pi^2)$.
The $q$-independent contribution here corrects the one erroneously given in
Eq. (55) of \cite{SPD05} (note the different sign conventions in
(\ref{Ie}) and Eq. (44) of \cite{SPD05}).
The precise value of this constant term was
redundant for the analysis carried out in that paper and hence did not influence
its results and conclusions.

Now, using Mathematica \cite{math} we obtain from (\ref{nn1})
\be\label{n2m2}
{\mathcal N}_2(m{=}2)=K_{3-\ve}K_2\frac{\ve}{c_{-1}}
\int_0^\infty{q\,dq\over(1+q^4)^2}\left[1{-}\ve i_0(q){+}O(\ve^2)\right]
=\ve-\ve^2\ln\frac{4}{3}+O(\ve^3).
\ee

Let us turn to the function $J(p,q)$ and the integral $\mathcal N_J$
(see (\ref{nn2}) and (\ref{JHG})--(\ref{J3})).
The first part of $J(p,q)$, $J^{(1)}(p,q)$,
is given by (\ref{J1}) with $h_{-1}=-1/(32\pi^2)$.
The function $J^{(2)}(1,q)$ is given by (\ref{JJ2}) where at $m=2$
the inner integral over $u$ is Gaussian. It is proportional to
$\e^{-q^2r}$. The angular integration in the three-dimensional $r$ integral
gives $(r/p)\sin pr$.
Numerical constants of these two integrations cancel giving
\be\label{J2}
J^{(2)}(p,q;m=2)={h_{-1}\over p}
\int_0^\infty dr\,\ln r\,\sin pr\,\e^{-q^2r}+O(\ve).
\ee
The remaining radial integral can be found, for example, in
\cite[865.911]{Dwight}. We obtain
\be\label{J22}
J^{(2)}(p,q)={h_{-1}\over p^2+q^4}\Big[\,{q^2\over p}\,\arctan{p\over q^2}
-\frac{1}{2}\ln\left(p^2+q^4\right)
-C_E\Big]+O(\ve).
\ee
In $J^{(3)}(p,q)$,
we perform the both angular integrations of (\ref{J3i}) explicitly and use,
along with (\ref{F2c}),
the closed form of the function $\Theta(u;m{=}2)$ known from \cite{SD01},
\be\label{Theta2}
\Theta(u;2)=-2\Big[C_E+\ln\frac{u^2}{4}+E_1\Big(\frac{u^2}{4}\Big)\Big]
=-2 \mbox{Ein}\Big(\frac{u^2}{4}\Big).
\ee
Here $E_1(x)$ is the exponential integral \cite{AS}
and $\mbox{Ein}(x)$ is the entire function
defined by the expression in square brackets \cite{AS}.
The radial $r$ integral in (\ref{J3i}) then reduces to
\be
\int_0^\infty dr \sin pr J_0(q u \sqrt r)=\frac{1}{p}\cos\frac{u^2 q^2}{4 p}
\ee
as can be found out from \cite[2.12.18.7]{PBM2}. Thus we remain with
\be
J^{(3)}(p,q;m=2)=-\frac{1}{32\pi^2}\frac{1}{p^2}
\int_0^\infty dx \e^{-x}\cos\frac{q^2 x}{p}\left[C_E+\ln x+E_1(x)\right]
\ee
where we changed the variable $u$ in favor of $x=u^2/4$.
It is convenient here to do the integration by parts
remembering that the derivative
of $E_1(x)$ is $-\e^{-x}/x$. The calculation can be completed with the help of
Mathematica \cite{math} yielding the result
\be\label{J33}
J^{(3)}(p,q)={h_{-1}\over p^2+q^4}\Big[\,{q^2\over p}\,\arctan{q^2\over 2p}
-{q^2\over p}\,\arctan{q^2\over p}
-\frac{1}{2}\ln\frac{p^2+q^4}{4 p^2+q^4}
\Big]+O(\ve).
\ee
Adding up all three contributions
(\ref{J1}), (\ref{J22}), and (\ref{J33})
to $J(p,q)$ and expressing the result in the form (\ref{Sj})
we derive the function $j_0(q)$:
\be\label{J0f}
j_0(q)=q^2\Big(\frac{\pi}{2}-2\arctan q^2+
\arctan\frac{q^2}{2}\Big)-\ln\frac{1+q^4}{\sqrt{4+q^4}}-\frac{1}{2}C_E
+\ln\sqrt\pi.
\ee
Finally, with the help of Mathematica \cite{math},
from (\ref{nn2}), (\ref{Ie}), and (\ref{Sj}) we obtain
\bea
{\mathcal N}_J(m=2)&=&-K_{3-\ve}K_2\frac{\ve^2}{c_{-1}}
\int_0^\infty{q\,dq\over(1+q^4)^2}\left[\frac{1}{\ve}+j_0(q)+2i_0(q)\right]
+O(\ve^3)\nonumber\\\label{njfin}
&=&-\ve+3\ve^2\ln\frac{4}{3}+O(\ve^3)\,.
\eea

Inserting into (\ref{nu2fin}) the known values \cite{SPD05}
of $\eta_{L2}$ and $\eta_{L4}$ at $m=2$ along with
the $\ve$ expansions (\ref{n2m2}) and (\ref{njfin}) for ${\mathcal N}_2$ and
${\mathcal N}_J$ we derive the correct result (\ref{R1n}) for
$\nu_{L2}^{-1}(m=2)$ at large $n$ and small $\ve$.

\section{A note on dimensional regularization and analytical continuation}
\label{d2}

Let us consider the validity region of our results.
These have been derived directly at LP
within a "massless" theory employing the dimensional regularization.
This is precisely the way of producing large-$n$ expansions accepted
in \cite{VPH81a}.
Above we have studied three situations where dimensional regularization
works well and leads to meaningful results. These included calculations
of critical exponents at CP, at isotropic LP,
and the epsilon expansion about the upper critical dimension for
$m$-axial anisotropic LP. The results in all these special cases
have been obtained before by other means,
but also mainly with the help of dimensional regularization.

As usual, the dimensional regularization implies a need in certain analytic continuations
(see e.g. \cite[Ch. 4]{Collins}).
In the theory of isotropic CPs it is possible to obtain most of results,
at least in lower-order approximations, in closed explicit forms.
Very often, albeit not necessarily, they contain Euler Gamma functions with negative arguments,
which require analytical continuation.
This is provided by the Cauchy-Saalsch\"utz formula
(see \cite[Ch. 12]{WhW}, \cite[Ch. 1]{ChZ})
\be\label{GAS}
\Gamma(\alpha)=\int_0^\infty dt t^{\alpha-1}\Bigg[\e^{-t}-\sum_{j=0}^k
\frac{(-t)^j}{j!}\Bigg]
\ee
where the Euler integral with subtractions converges for $-k-1<\Re\,\alpha<-k$.
In practice, the relation $\Gamma(\alpha)=\Gamma(\alpha+1)/\alpha$ for $-1<\alpha<0$
is (repeatedly) used, which gives meaning to such Gamma functions.
For example, the coefficient $\eta^{(1)}$ 
from Eq. (28) of \cite{VPH81a} contains the function $a(2)=\Gamma(d/2-2)$
where $d/2-2<0$ for $d<4$. The last procedure gives here
$\Gamma(d/2-1)$ where the argument is already positive for physical range $d>2$.

Accepting the approach of \cite{VPH81a}
we worked directly at LP and used the dimensional regularization.
This implied vanishing of coefficients at $\phi^2$ and $(\nabla_z\phi)^2$
in the Hamiltonian (see (\ref{EHI}), (\ref{EH})).
No subtractions like that in (\ref{GAS}) appeared in the
resulting integrals, and it was possible to use the same transformation
in isotropic limits and for infinitesimally small values of $\ve$.

However, generic results contain integrals normally not expressible in a closed form.
We have already had a hint that some analytic continuations will be needed
beyond the special cases just discussed.
Indeed, working in the $\ve$ expansion we saw
that the integral $J(p,q)$ from (\ref{J1q}) exhibits an infrared $1/\ve$ pole.
However, while the dimensional
regularization was employed in constructing the $\ve$ expansion, there was no
problem in combining it with the ultraviolet $1/\ve$ pole of $I(p,q)$ and
arriving at correct final results for $d\lesssim d^*$.

As usual, infrared problems become more severe when one moves towards
lower space dimensions. For example, it is not possible to
calculate numerically ${\mathcal N}_J$
directly at $d=4$ or $3$ and $m=1$
due to infrared divergence of the integral $J(1,q)$.
In order to obtain sensible results
the definition (\ref{J1q}) of the function $J(p,q)$
has to be extended so that it becomes well-defined
in the whole region of interest, that is in the interval $0<\ve<2$.

Realizing that infrared problems are typical for massless theories
we have done \cite{SD11} another calculation of
$\gamma_L$ employing the massive theory.
We used the method of Ma \cite{Ma73} where
in constructing the large-$n$ expansion the calculations are carried out
not directly at the transition point, but away from it.
An asymptotic analysis of the theory is performed
for infinitesimally small but not vanishing "mass",
the role of which plays the inverse susceptibility.
In this way we obtained the results
for LP's critical exponents of the same general form as above.
An essential modification occurs only in the integral representation of
the function $J(p,q)$.
The massive theory generates appropriate subtractions, similar to that
of (\ref{GAS}), in the integrand of $J(p,q)$.
This provides the analytic continuation for this function
of the type discussed in \cite[Ch. 4.2]{Collins}.
In the region $0<\ve<1$ it becomes, instead of (\ref{J1q}),
\be\label{Jsub}
J^{(1{\rm s})}(p,q)=\int_{\bm p'}^{(d-m)}\int_{\bm q'}^{(m)}{1\over (p'^2+q'^4)^2}
\left[ {1\over (\bm{p}'+\bm p)^2+(\bm {q}'+\bm q)^4}-\frac{1}{p^2+q^4}\right].
\ee
For larger values of $\ve$, the subtraction of $1/(p^2+q^4)$
is not sufficient and the integral (\ref{Jsub})
again diverges at small integration momenta.
In the complementary region $1<\ve<2$ an additional
subtraction appears, and we must use the formula
\be\label{Jds}
J^{(2{\rm s})}(p,q){=}\int_{\bm p'}^{(d-m)}\int_{\bm q'}^{(m)}
{1\over (p'^2{+}q'^4)^2}
\left[ {1\over (\bm{p}'{+}\bm p)^2+(\bm {q}'{+}\bm q)^4}-\frac{1}{p^2{+}q^4}
-q'^{\,2}\,S_2(p,q)\right]
\ee
where the function $S_2(p,q)$ is defined by
\be\label{S2fun}
S_2(p,q)=\frac{2}{m}{q^2\over(p^2+q^4)^3}\left[(6-m)q^4-(2+m)p^2\right].
\ee
Conversely, this doubly subtracted form cannot be used for $\ve<1$ since there
the last term of (\ref{Jds}) will cause an ultraviolet divergence.

Thus, for non-infinitesimal values of $\ve$ from the range
$0<\ve<2$, the function $J(p,q)$ has to be replaced in
(\ref{nn2})--(\ref{Cn4}) by its appropriate
"subtracted" version $J^{(1{\rm s})}(p,q)$ or $J^{(2{\rm
s})}(p,q)$ from (\ref{Jsub}) or (\ref{Jds}). Hence follow the
results for critical exponents $\nu_{L2}$, $\nu_{L4}$ and
$\gamma_L$, which are valid, as in \cite{SPD05}, for generic
$m$-axial LPs with $0<m<d$ in the whole stripe
$2+m/2<d<4+m/2$ between the lines of the lower and upper critical
dimensions.

Using the analytic continuation of $J(1,q)$
defined by (\ref{Jds}) and (\ref{S2fun})
we have calculated the $1/n$ coefficients for uniaxial systems ($m=1$)
directly at $d=4$ and $d=3$ where $\ve=1/2$ and $\ve=3/2$, respectively.
These results will be discussed in the following sections.

\section{Special case $\bm{d=4}$, $\bm{m=1}$}\label{d4}

In this section we consider the special case of four-dimensional systems
with uniaxial anisotropy. As discussed in the Introduction, this can be related
to quantum field theories with violation of the Lorentz invariance.
Fortunately, when $d=4$ and $m=1$ all calculations
for critical indices to order $O(1/n)$ can be done analytically.
Details of calculations and results can be compared with that
of Anselmi \cite{Ans08}.

Here we have to use (\ref{Jsub}) for the function $J(1,q)$
in the integrand of $\mathcal N_J$ in (\ref{nn2}).
The integral over $p'$ is now three-dimensional
and it involves
\be
\int_{\bm p'}^{(3)}{1\over (p'^2+q'^4)^2}
{1\over (\bm{p}'+\bm p)^2+(\bm q'+\bm q)^4}=
\frac{1}{8\pi}\,\frac{1}{q'^2}\,{1\over\left[q'^2+(\bm q'+\bm q)^2\right]^2+p^2}.
\ee
For (\ref{Jsub}) we get
\be\label{JQ!}
J^{(1{\rm s})}(p,q)=\frac{1}{16\pi^2}\int_{-\infty}^\infty
\frac{dx}{x^2}
\left\{{1\over\left[x^2+(x+q)^2\right]^2+p^2}-\frac{1}{p^2+q^4}\right\}.
\ee
The subtraction in curly brackets provides the convergence
at $x\to0$. Simple algebraic transformations reduce the last integral to
\be\label{JQ@}
J^{(1{\rm s})}(p,q)=\frac{1}{\pi^2}
\lrs{q^2\,\frac{q^4-p^2}{q^4+p^2}\,J_0(p,q)
-\frac{1}{2}\,J_2(p,q)-\frac{2}{q^4+p^2}\,J_6(p,q)}
\ee
where $J_k(p,q)$ with $k=0,2,6$ are given by
\be
J_k(p,q)=\int_0^\infty\frac{x^kdx}{A_-A_+}\quad\mm{with}\quad
A_\pm\equiv{1\over\left[x^2+(x\pm q)^2\right]^2+p^2}.
\ee
The integrals $J_k(p,q)$ can be calculated with the help of the residue calculus.
For instance,
\be
J_0(p,q)=\frac{3\pi}{16\sqrt 2}\;p^{-1}\,\frac{q^4+q^2\sqrt{4p^2+q^4}-2p^2}
{(2q^4-p^2)(p^2+q^4)\sqrt{4p^2+q^4}}\sqrt{\sqrt{4p^2+q^4}-q^2}\,.
\ee
For the whole combination (\ref{JQ@}) the result is
$J^{(1{\rm s})}(p,q)=p^{-5/2}J^{(1{\rm s})}(1,q)$ with
\be\label{J!S}
J^{(1{\rm s})}(1,q)=\frac{1}{8\sqrt 2\;\pi}\,
\frac{(q^4-1)\sqrt{\sqrt{4+q^4}+q^2}-2q^2\sqrt{\sqrt{4+q^4}-q^2}}
{(1+q^4)^2\sqrt{4+q^4}}\,.
\ee
This function is recognized in the second term
of \cite[(C.2)]{Ans08} after removing the auxiliary mass $m^2$
introduced there artificially to avoid IR problems in evaluating the
diagram called the scalar triangle.

As an illustration to the discussion of Section \ref{d2}, in Appendix \ref{app3}
we rederive the result (\ref{J!S}) in coordinate representation
without any subtraction like that of (\ref{Jsub}).
There, the required analytic continuation is performed by other means and
involves the relation $\Gamma(x)=\Gamma(1+x)/x$ for a negative $x$.
We also believe that it is worth to give detailed calculations
of different kinds for four-dimensional systems at LP in view
of growing interest to such systems in quantum field theory.

Using $I^{-1}(1,q)=4\pi\sqrt 2\sqrt{\sqrt{4+q^4}+q^2}$ from \cite{SPD05}
along with (\ref{J!S}) we get
\be
\mathcal N_2=\frac{2}{\pi\sqrt 3}\quad\mm{and}\quad \mathcal
N_J=-\frac{7}{24\pi\sqrt 3}\,.
\ee
These agree with the terms (b) and (c) from \cite[(8.3)]{Ans08}.
Further, with $\eta_{L2}^{(1)}=5/(9\pi\sqrt 3)$
and $\theta^{(1)}=-4/(27\pi\sqrt 3)$ derived in \cite{SPD05},
we obtain from (\ref{Z1fin}) and (\ref{nu2fin})
\be\label{ZNU}
\zeta_1=-\frac{103}{18\pi\sqrt 3}\mmm{and}
\nu_{L2}^{-1}=\frac{3}{2}+\frac{301}{54\pi\sqrt 3}\;\frac{1}{n}+O(n^{-2})
\ee
for four-dimensional uniaxial systems.

Both $\zeta_1/n$ and the $O(1/n)$ contribution in $\nu_{L2}^{-1}$ from (\ref{ZNU})
do not match the value $\gamma_\sigma=-83/(9\pi n\sqrt 3)$ from
\cite[(8.4)]{Ans08}. Such value results from the combination
$\gamma_\sigma=2[4\mathcal N_J-2\mathcal N_2+\eta_{L2}^{(1)}]$.
Up to the sign of $\eta_{L2}^{(1)}$, it is similar to $2\zeta_1/n$
given by (\ref{Z1fin}), but has no direct counterpart in our theory.

\section{Numerical results for uniaxial systems in $\bm{d=3}$
}\label{d3}

The choice $m=1$ and $d=3$,
corresponds to the experimentally accessible case of
three-dimensional systems with uniaxial anisotropy.
Here we have to use numerical means to evaluate the $1/n$ coefficients
of the exponents $\nu_{L2}^{-1}$, $\gamma_{L}$ and $\nu_{L4}$,
given by (\ref{nu2fin}), (\ref{Gam})--(\ref{Gamc}) and
(\ref{Fn4})--(\ref{Cn4}), respectively.

The integral ${\mathcal N}_2(m{=}1,d{=}3)$ (see (\ref{nn1})) involves the
function $I(1,q)$ known explicitly from \cite{SPD05},
\be\label{SDD}
I(1,q)=\frac{1}{2\pi}(1+q^4)^{-1/2}(4+q^4)^{-1/4}\bm K(k),
\ee
where $\bm K(k)$ is the complete elliptic integral of the first kind
\cite[Ch. 13]{BE3}, and
\be\label{SDT}
k^2=\frac{1}{2}\left(1-{q^2\over\sqrt{4+q^4}}\,{3+q^4\over 1+q^4} \right),
\qquad 0\le k^2\le1/2.
\ee
The numerical result is ${\mathcal N}_2(m{=}1,d{=}3)\simeq 0.249788$.

In the calculation of ${\mathcal N}_J(m{=}1,d{=}3)$ (see (\ref{nn2}))
we have to use, in addition,
the improper integral $J^{(2{\rm s})}(1,q)$ from
(\ref{Jds})--(\ref{S2fun}), at $\ve=3/2$.
Numerically, using FORTRAN, we obtain ${\mathcal N}_J(m{=}1,d{=}3)\simeq 0.1096$.
In fact, an involved calculation in the complex plane
yields the explicit expression
\bea\label{J23}
J^{(2{\rm s})}(1,q)&=&\frac{1}{4\pi}(1+q^4)^{-7/2}(4+q^4)^{-3/4}\left\{
q^2(9-46q^4+7q^8)\Big[2\bm E(k)-\bm K(k)\Big]\right.
\nonumber\\
&+&\left.(1+q^4)(1+7q^4)\sqrt{4+q^4}\;\bm K(k)\right\}
\eea
where $\bm E(k)$ is the complete elliptic integral of the second kind.
The both functions $\bm K(k)$ and $\bm E(k)$ are related to
Gauss hypergeometric functions with the argument (\ref{SDT})
via \cite[Ch. 13]{BE3}, \cite[Ch. II.16]{PBM3}
\be
\bm K(k)=\frac{\pi}{2}\,_2F_1\Big({\frac{1}{2}},\frac{1}{2};1;k^2\Big),
\qquad
\bm E(k)=\frac{\pi}{2}\,_2F_1\Big({-\frac{1}{2}},\frac{1}{2};1;k^2\Big),
\quad |k|<1.
\ee
With the knowledge of (\ref{J23}) we checked the numerical value
of ${\mathcal N}_J(m{=}1,d{=}3)$ using Mathematica \cite{math}.

Using the coefficients $\eta_{L2}^{(1)}(1,3)\simeq 0.306$,
$\theta^{(1)}(1,3)\simeq -0.0487$ and $\eta_{L4}^{(1)}(1,3)\simeq 0.223$
from \cite{SPD05} along with above values of ${\mathcal N}_2$ and ${\mathcal N}_J$,
we obtain
\bea
&&\nu_{L2}(1,3)=2-1.274/n+O(n^{-2}),\;\quad
\nu_{L2}^{-1}(1,3)=\frac{1}{2}+0.319/n+O(n^{-2})
\nonumber\\&&
\gamma_L(1,3)=4-3.161/n+O(n^{-2}),\;\;\quad
\gamma_L^{-1}(1,3)=\frac{1}{4}+0.198/n+O(n^{-2})
\nonumber\\&&
\nu_{L4}(1,3)=1-0.734/n+O(n^{-2}),\;\quad
\nu_{L4}^{-1}(1,3)=1+0.734/n+O(n^{-2})
\nonumber\\&&
\alpha_L(1,3){=}{-}3+3.283/n{+}O(n^{-2}),\;\;\,\quad
\alpha_L^{-1}(1,3){=}{-}\frac{1}{3}-0.365/n{+}O(n^{-2})
\nonumber\\\nonumber&&
\beta_L(1,3)=\frac{1}{2}-0.0612/n{+}O(n^{-2}),\;\quad
\beta_L^{-1}(1,3)=2+0.245/n{+}O(n^{-2}).
\eea

Comparing these $1/n$ expansions with corresponding $\ve$ expansions
\cite{SD01} we see that already
zeroth-order terms of both, except for $\beta_L$, are quite different.
Indeed, the $\ve$ expansions of $\nu_{L2}$, $\gamma_L$, $\nu_{L4}$, $\alpha_L$
and $\beta_L$ start with
mean-field values $1/2,\,1,\,1/4,\,0,\,1/2$, respectively.
These do not depend on the number of anisotropy axes $m$ and actually coincide
with mean-field values of usual CP exponents.
By contrast, the large-$n$ limits of these exponents are built on the basis
of the combination $d-m/2-2=2-\ve$,
twice the classical dimension of the field $\phi(x)$.
This value essentially depends both on $m$ and $d$.
Otherwise, the values of coefficients
of the $1/n$ corrections are quite reasonable.
Except for $\alpha_L$ and $\alpha_L^{-1}$,
they are never larger compared to the corresponding zeroth-order terms.
Hence, there is no hope that numerical estimates derived from
the short series expansions in $1/n$ will be in tight agreement
with that from the $\ve$ expansion.
Especially this refers to the exponents whose zeroth orders strongly differ
in both these approaches. This is anyway not surprising. It is well-known
that the truncated large-$n$ expansions do not give
very good numerical estimates.

For example, what we can get from the large-$n$ expansions of
$\nu_{L2}(1,3)$ and $\nu_{L2}^{-1}(1,3)$ at $n=3$, is
$\nu_{L2}(1,3;3)\simeq 1.6$ as compared to the second-order
$\ve$ expansion's value \cite{SD01} $0.8$.
In the case of $\beta_L$, where the both types of series have
equal leading terms, we obtain $\beta_L(1,3;3)\simeq 0.48$,
while the second-order $\ve$ expansion gives the value 0.3.

\section{Concluding remarks}

The present article represents an extension of our previous work \cite{SPD05}
on the large-$n$ expansion for $m$-axial LPs.
In addition to the $O(1/n)$ values of the correlation exponents
$\eta_{L2}$ and $\eta_{L4}$ calculated in
that reference, here we derived the non-trivial $1/n$ corrections of the
thermal exponents $\nu_{L2}$, $\nu_{L4}$, $\gamma_L$, $\alpha_L$, and $\beta_L$.
We did it by using two different approaches: (i) the technique of matching the asymptotic
scaling forms of full propagators \cite{VPH81a,Vas98}, and (ii) the method of
Ma \cite{Ma73} of the asymptotic analysis of the correlation function at small but non-vanishing
temperature deviations from the LP.

In sections \ref{S1} and \ref{S2} we described in full detail
the calculations employing the method (i).
This yields the result (\ref{nu2fin}) for the perpendicular correlation-length exponent
$\nu_{L2}$, and further exponents have been obtained through the scaling laws.
Starting from (\ref{nu2fin}) and using the dimensional regularization it was possible to
show that: \emph{(a)} at $m=0$ we correctly recover the large-$n$ results of the
usual CP theory, \emph{(b)} the known large-$n$ results for isotropic LPs are
reproduced at $m=d$, \emph{(c)} an explicit calculation at $m=2$
yields the correct expressions of the epsilon expansion up to $O(\ve^2/n)$,
\emph{(d)} the compatibility with the epsilon-expansion results for generic $m$
can be established along the lines of \cite{SDP08}.

However, it was still impossible to use the formula (\ref{nu2fin}) as it stands
to compute the values of $1/n$ corrections for lower dimensions $d$
and $m\ne\{0,d\}$, which should correspond to different cases of anisotropic systems.
Employing the method (ii) yielded subtractions
in the integral representation of the function $J(p,q)$, defined in (\ref{J1q}),
necessary for its analytic
continuation to arbitrary dimensions $d$ and $m$ of the main interest: These
lie in the $(m,d)$ plane between the lines $d=d^*(m)=4+m/2$, $d=d_\ell(m)=2+m/2$,
$m=0$, and $m=d$. In this region of $d$ and $m$,
the relevant expressions are (\ref{Jsub}) and (\ref{Jds}) along with (\ref{S2fun}).
These, instead of (\ref{J1q}), have to be used in the integrand of
${\mathcal N}_J$ in (\ref{nn2}) which, in turn, has to be inserted into (\ref{nu2fin}).

Proceeding in this way we performed explicit calculations in the special case
$d=4$, $m=1$ considered by Anselmi \cite{Ans08} in the context of the Lorentz violating
field theory. We compared our calculations and results with that of \cite{Ans08}
in Sec. \ref{d4}.

At last, in the case of three-dimensional systems with uniaxial anisotropy ($d=3$, $m=1$),
which are of main interest for condensed-matter physics, we were able to get
our final results for critical exponents in the numerical form.
These are compared with the estimates stemming from the epsilon expansion.

\ack
The authors are grateful to H. W. Diehl for cooperation at the initial stage
of this work and for extended comments on the manuscript.
Yu.M.P. is  grateful to L. Tz. Adzhemyan, J. Honkonen, and
A. N. Vassiliev for their valuable help and discussions.
M.A.Sh and Yu.M.P thank H. W. Diehl and
Fakult\"at f\"ur Physik for their hospitality at the University of Duisburg-Essen.

\appendix \section{Epsilon expansion of $\bm{I(p,q)}$}\label{app1}

It is convenient to perform the $\ve$ expansions of the integrals $I(p,q)$
and $J(p,q)$ in direct space.
We introduce, following \cite{DS00,SD01}, the Fourier transformation
of the free propagator (\ref{Sphl}) via
\be\label{Gx}
G^{(0)}(x)\equiv G^{(0)}(r,z)=\int_{\bm p}^{(d-m)}
\int_{\bm q}^{(m)}{\e^{i\bm{p}\bm r}\e^{i\bm {q}\bm z}\over p^2+q^4}=
r^{-2+\ve}\,\Phi(u),
\qquad u=\frac{z}{\sqrt r}.
\ee
The scaling function $\Phi(u)\equiv\Phi(u;m,\ve)$
is defined by the momentum integral
from (\ref{Gx}) with $\bm r\to\bm 1$ and $\bm z\to\bm u$.
Its mathematical properties are discussed in detail in \cite{SD01}.

In the coordinate representation, the integral $I(p,q)$ defined in (\ref{Ipq})
is given by
\be
I(p,q)=\int d^dx [G^{(0)}(x)]^2\, \e^{-i\bm p\bm r}\e^{-i\bm q\bm z}\,
\mm{where} \int d^dx\equiv \int d^{d-m}r\int d^mz.
\ee
Adding and subtracting unity at $\e^{-i\bm q\bm z}$ and taking $p=1$ we
split the resulting integral via $I(1,q)=I(1,0)+I_0(1,q)$ into a constant part
\be
I(1,0)=\int d^dx [G^{(0)}(x)]^2\, \e^{-i\bm 1\bm r}
\ee
and the function $I_0(1,q)$ vanishing at $q=0$ and finite at $d=d^*(m)$:
\be\label{I1pq}
I_0(1,q)=\int d^dx [G^{(0)}(x)]^2\,\e^{-i\bm1\bm r}
\left(\e^{-i\bm q\bm z}-1\right).
\ee
The constant $I(1,0)\equiv F_{m,\ve}/\ve$ (see \cite{SD01})
has been calculated for arbitrary $m$ before \cite{DS00,SD01,SPD05},
\be\label{Coni}
I(1,0)=(4 \pi)^{-{d\over 2}}\,{1\over 2}\,\Gamma\Big({\ve\over 2}\Big)\,
{\left[\Gamma(1-{\ve\over 2})\right]^2\over \Gamma(2-\ve)}\;
{\Gamma({m\over 4})\over \Gamma({m\over 2})}=
\frac{c_{-1}}{\ve}\lrs{1+c_0\ve+O(\ve^2)}\,.
\ee
The $m$-dependent pole coefficient $c_{-1}\equiv F_{m,0}$
appears in Eq. (\ref{Ie}).
The finite-part constant $c_0=(2-C_E+\ln4\pi)/2$ is the same for any $m$.
We cannot calculate the function $I_0(1,q)$
in closed form for arbitrary $m$. Nevertheless, a formal $\ve$ expansion
\be\label{IFO}
I(1,q)=\frac{c_{-1}}{\ve}
\Big\{1+\big[c_0+c_{-1}^{-1}I_0(1,q)\big]\,\ve+O(\ve^2)\Big\}
\ee
appears to be useful in our calculations of Section \ref{APC}.

\section{Epsilon expansion of $\bm{J(p,q)}$}\label{Lex}

To calculate $J(p,q)$, we introduce the Fourier transformation
of $(p^2+q^4)^{-2}$ via
\be\label{Hx}
H(x)\equiv H(r,z)=\int_{\bm p}^{(d-m)} \int_{\bm q}^{(m)}
{\e^{i\bm{p}\bm r}\e^{i\bm {q}\bm z}\over(p^2+q^4)^2}=
r^\ve\,\Psi(u)\,.
\ee
In direct space, the function $H(x)$ represents a
convolution of two free propagators, a quite complicated mathematical object.
From \cite{SD01} we know that the scaling function $\Psi(u)$ has the Taylor
series expansion
\be\label{HPS}
\Psi(u)=b_m\pi^{\ve/2}\sum_{k\ge 0}{1\over k!}\,
{\Gamma(-{\ve\over 2}+\frac{k}{2})\over
\Gamma({1\over 2}+{m\over 4}+\frac{k}{2})}\Big(-{u^2\over 4} \Big)^k
\ee
with $b_m=2^{-4-m}\pi^{-(6+m)/4}$.
The Laurent expansion of $\Psi(u)$ is implemented by
splitting out the $k=0$ term of the sum.
At small $\ve$, it contains an $1/\ve$ pole
with the residuum $h_{-1}=-2b_m/\Gamma(1/2+m/4)$.
For arbitrary $m$ we have $h_{-1}=-c_{-1}$ where $c_{-1}$ is the pole coefficient
from (\ref{Coni}). The remaining series with $k\ge1$ is of order $O(1)$.

Hence, the function $H(x)$ from (\ref{Hx}) can be written as
$H(x)=H(r,0)+H_0(r,z)$ where
\bea\label{H0H}
&&H(r,0)=r^\ve\,\Psi(0)=
h_{-1}\Big({1\over\ve}+C_E/2+\ln\sqrt\pi+\ln r\Big)+O(\ve),\\&&
\label{H1H}
H_0(r,z)=r^\ve\lrs{\Psi(u)-\Psi(0)}=b_m\,\Theta(u;m)+O(\ve).
\eea
The function $\Theta(u;m)$ is given by\footnote{This is
the correct version of the misprinted formula (D5) of \cite{SD01}.}
\be\label{TUM}
\Theta(u;m)=\sum_{k\ge 1}{1\over k!}\,{\Gamma(\frac{k}{2})\over
\Gamma({1\over 2}+{m\over 4}+\frac{k}{2})}\Big(-{u^2\over 4} \Big)^k
=\lim_{\ve\to 0}\Big[B^{-1}(\ve)\Psi(u)-
\frac{\Gamma(-\frac{\ve}{2})}{\Gamma(\frac{1}{2}+\frac{m}{4})}\Big].
\ee
On the other hand, we can handle the integral from (\ref{Hx})
as in Appendix \ref{app1}, by adding and subtracting $1$ at $\e^{i \bm q\bm z}$.
This gives
\be\label{HR}
H(r,0)=\int_{\bm p}^{(d-m)} \int_{\bm q}^{(m)}{\e^{i\bm p\bm r}\over(p^2+q^4)^2}
\mm{and}
H_0(r,z)=\int_{\bm p}^{(d-m)}\e^{i\bm p\bm r}\,\int_{\bm q}^{(m)}
{\e^{i\bm q\bm z}-1\over(p^2+q^4)^2}.
\ee

Let us return to the function $J(p,q)$. In the coordinate representation it reads
\be\label{JHG}
J(p,q)=\int d^dx H(x)G^{(0)}(x)\, \e^{-i\bm{p}\bm r}\e^{-i\bm {q}\bm z}\,.
\ee
In order to proceed we use here the $\ve$ expansion
of $H(x)$ given by (\ref{H0H})--(\ref{H1H}).
This yields three different terms contributing to
$J(p,q)$:
\be\label{J3}
J(p,q)=J^{(1)}(p,q)+J^{(2)}(p,q)+J^{(3)}(p,q).
\ee

The first of them, involving only the constant contributions
from $H(x)$, is trivial because the integration over $x$ in (\ref{JHG}) gives
in this case simply the inverse Fourier transformation for $G^{(0)}(x)$.
Thus, by (\ref{Gx}) and (\ref{H0H}) we obtain
\be\label{J1}
J^{(1)}(p,q)=h_{-1}\Big({1\over\ve}+C_E/2+\ln\sqrt\pi\Big)
{1\over p^2+q^4}+O(\ve)\,,
\ee
valid for arbitrary $m$.
This is the only contribution to $J(p,q)$ that contains a pole term.
The remaining parts of $J$ can be evaluated directly
at the upper critical dimension.

Taking the $\ln r$ term from (\ref{H0H}) we have
\be\label{JJ2}
J^{(2)}(p,q)=h_{-1}\int\!\! d^{d^*-m}r \; r^{-2+m/2}\,
\ln r\;\e^{-i\bm{p}\bm r}
\int\!\! d^m u \; \Phi(u)\e^{-i\bm {q}\bm u\sqrt{r}}+O(\ve)\;.
\ee
Here, the inner integral over $u$ represents a Fourier transformation inverse
to that given in Eq. (14) of \cite{DS00}, which is a direct consequence
of the definition (\ref{Gx}): We have
\be
\int\!\! d^m u \; \Phi(u)\e^{-i\bm {q}\bm u\sqrt{r}}=
\frac{(q^2r)^{\mu}}{(2 \pi)^{\mu+1}}\,K_{\mu}(q^2r)
\mm{with}\mu\equiv\frac{d{-}m}{2}{-}1=1{-}\frac{m}{4}{-}\frac{\ve}{2}.
\ee
Hence, after angular integration in the $\bm r$ subspace
we can write $J^{(2)}(p,q)$ as
\be
J^{(2)}(p,q)=h_{-1}\,p^{-\mu}q^{2\mu}\int_0^\infty dr\,r\,\ln r\;J_{\mu}(pr)\,
K_{\mu}(q^2r)+O(\ve)\,.
\ee
In the last two equations $J_{\mu}(x)$ and $K_{\mu}(x)$ are
the Bessel functions of the first and second kind \cite{AS}, respectively.
Noting that the logarithm in the integrand can be represented as
$\lim_{\alpha\to 0}(r^\alpha-1)/\alpha$ we can do the last integral
by using \cite[2.16.21.1]{PBM2}.
This reference tells us that the result
for the integral $J_\alpha^{(2)}(p,q)$
involving $r^\alpha$ is proportional to the
Gauss hypergeometric function $_2F_1(1+\alpha/2,\mu+1+\alpha/2;\mu+1;-p^2/q^4)$.
Applying to this function the linear transformation 7.3.1.3 from \cite{PBM3}
and taking $\ve=0$ we get
\bea
J_\alpha^{(2)}(p,q)&=&h_{-1}2^\alpha\Gamma\Big(1+\frac{\alpha}{2}\Big)
\frac{\Gamma\Big(2-\frac{m}{4}+\frac{\alpha}{2}\Big)}
{\Gamma\Big(2-\frac{m}{4}\Big)}
\nonumber\\&\times&
(p^2+q^4)^{-1-\frac{\alpha}{2}}\,
_2F_1\Big(-\frac{\alpha}{2},1+\frac{\alpha}{2};
2-\frac{m}{4};\frac{p^2}{p^2+q^4}\Big).
\eea
Since one of the nominator parameters of the function $_2F_1$ is
proportional to $\alpha$ it is trivial to expand this function to
first order in $\alpha$. By a straightforward calculation we obtain
\bea\label{JR2}
J^{(2)}(p,q)&=&\lim_{\alpha\to 0}\frac{1}{\alpha}
\lrs{J_\alpha^{(2)}(p,q)-J_0^{(2)}(p,q)}
=\frac{h_{-1}}{p^2+q^4}\left[\frac{1}{2}\right.\psi\Big(2-\frac{m}{4}\Big)
-\frac{1}{2}C_E+\ln2
\nonumber\\
&-&\left.\frac{1}{2}\ln(p^2+q^4)-\frac{2}{8{-}m}\,\frac{p^2}{p^2{+}q^4}\;
_2F_1\Big(1,1;3-\frac{m}{4};\frac{p^2}{p^2+q^4}\Big)\right]+O(\ve).
\eea

The last contribution to $J(p,q)$ implied by (\ref{H1H}) and (\ref{J3}) is
\be\label{J3i}
J^{(3)}(p,q)=b_m\int d^dx\,\Theta(u;m) G^{(0)}(x)\,\e^{-i\bm{p}\bm r}
\e^{-i\bm {q}\bm z}+O(\ve)\,.
\ee
It can be left in the present state for the calculation with arbitrary $m$,
and in the special case $m=2$, $d^*=5$ we give an explicit expression
for $J^{(3)}(p,q)$ in (\ref{J33}).

\section{$\bm{J(p,q)}$ at $\bm{d=4}$ and $\bm{m=1}$}\label{app3}

We start from the integral representation (\ref{JHG}) for $J(p,q)$.
As in \cite{SPD05}, we use mathematical simplifications
occurring on the line $d=m+3$, to which belongs the point $d=4$, $m=1$.
For $d=m+3$ we have
\be\label{C1}
\Phi_{d=m+3}(u)=(4\pi)^{-2+\ve}\e^{-x}\;\mbox{and}\;
\Psi_{d=m+3}(u)=2 b_m\pi^{\ve/2}x^\ve\lrs{\Gamma(-\ve)-\Gamma(-\ve,x)}
\ee
with $x\equiv u^2/4$. The expression for $\Psi(u)$ follows
from its general definition (\ref{HPS}).
$\Gamma(-\ve,x)$ is the complementary incomplete Gamma function
defined by the integral (see \cite{BE2})
\be
\Gamma(a,z)=\int_z^\infty t^{a-1}\e^{-t}dt
\ee
that converges for arbitrary $a$ when $\Re z>0$.
The Euler Gamma function $\Gamma(-\ve)$ is defined through
the analytic continuation of the same integral at $z=0$ (see (\ref{GAS})).

Using the scaling forms (\ref{Gx}) and (\ref{Hx}) for the functions
$G^{(0)}(x)$ and $H(x)$ involved
in (\ref{JHG}) and performing there angular integrations in both the
three- and $m$-dimensional sub-integrations over $r$ and $u$ we get
\be\label{Cw3}
J(p,q)=\frac{q^\ve}{p}2^{3-\ve}\pi^{2-\ve}\int_0^\infty du u^{1-\ve}
\Phi(u)\Psi(u)\int_0^\infty dr r^{3\ve/2}\sin(pr)J_{-\ve}(qu\sqrt r).
\ee
This integral representation is valid
for the whole line $d=m+3$ with $0\le m\le 2$ and $\ve$ in the same limits.
In the special case $d=4$, $m=1$ we have to
take here and in the scaling functions $\Phi$ and $\Psi$
from (\ref{C1}) $\ve=1/2$.
At $\ve=1/2$, the Bessel function $J_{-\ve}(z)\sim \cos z/\sqrt z$
and the inner integral over $r$ does not converge.
Nevertheless, it can be treated
in the spirit of dimensional regularization in order to assign to it a
sensible value. Using \cite[2.12.18.4]{PBM2},
the $r$ integral from (\ref{Cw3}) can be written down explicitly.
The result is a linear combination of two generalized hypergeometric
functions $_2F_3$. This can be evaluated at $\ve=1/2$ yielding
the needed analytical continuation for the last integral over $r$.
A simpler way to do the same thing is to use the formal relation
\bea
&&\int_0^\infty dr r^{1/2}\sin(pr)\cos(qu\sqrt r)
:=-2\frac{\partial}{\partial p}\int_0^\infty dx \cos(px^2)\cos(qux)
\nonumber\\&&\quad
=\frac{1}{4}\sqrt\frac{\pi}{2}\;p^{-5/2}\,(qu)^{-1/2}
\Big[(2p+q^2u^2)\cos\frac{q^2u^2}{4p}
+(2p-q^2u^2)\sin\frac{q^2u^2}{4p}\Big]
\eea
of the needed integral to the well-defined (for $p$, $qu>0$)
integral from \cite[2.5.22.5]{PBM1} expressed in terms of simple
trigonometric functions. Using this finding we can write
\be
J(p,q)=p^{-5/2}\,(2\pi)^{3/2}\lrs{j(w)+2wj'(w)}
\ee
where, with $w\equiv q^2/p$,
\bea
j(w)&=&\int_0^\infty du \Phi(u)\Psi(u)
\Big(\cos\frac{wu^2}{4}+\sin\frac{wu^2}{4}\Big)
\nonumber\\&&\strut=
2^{-7}\pi^{-3}\int_0^\infty dx \e^{-x}\lrs{\Gamma(-1/2)-\Gamma(-1/2,x)}
(\cos wx+\sin wx).
\eea
The last integral can be done using integration by parts
taking into account that $\Gamma'(-1/2,x)=-x^{-3/2}\e^{-x}$.
A straightforward calculation yields, with the help of
Mathematica \cite{math}, the result (\ref{J!S}) obtained in the main text
using the momentum representation.

\end{document}